\documentclass[preprint]{aastex}

\usepackage{rotating} 
\usepackage{amssymb}
\usepackage{subfigure}
\usepackage{bm}
\usepackage[normalem]{ulem}

\usepackage{graphicx}
\usepackage{epstopdf}
\usepackage[fleqn]{amsmath} 

\def\mean#1{\left< #1 \right>}

\allowdisplaybreaks

\begin{document}
\title{The transfer of resonance line polarization with partial frequency 
redistribution in the general Hanle-Zeeman regime}

\author{E. Alsina Ballester\altaffilmark{1,2}, 
L. Belluzzi\altaffilmark{3,4}, and 
J. Trujillo Bueno\altaffilmark{1,2,5}} 

\altaffiltext{1}{Instituto de Astrof\'{\i}sica de Canarias, E-38205 La 
Laguna, Tenerife, Spain}
\altaffiltext{2}{Departamento de Astrof\'{\i}sica, Facultad de F\'{\i}sica, 
Universidad de La Laguna, Tenerife, Spain}
\altaffiltext{3}{Istituto Ricerche Solari Locarno, CH-6605 Locarno Monti, 
Switzerland}
\altaffiltext{4}{Kiepenheuer-Institut f\"ur Sonnenphysik, D-79104 
Freiburg, Germany}
\altaffiltext{5}{Consejo Superior de Investigaciones Cient\'{\i}ficas, Spain}

\email{ealsina@iac.es}

\begin{abstract}
The spectral line polarization encodes a wealth of information about the 
thermal and magnetic properties of the solar atmosphere.
Modeling the Stokes profiles of strong resonance lines is, however, a complex 
problem both from the theoretical and computational point of view, especially 
when partial frequency redistribution (PRD) effects need to be taken into 
account.
In this work, we consider a two-level atom in the presence of magnetic fields 
of arbitrary intensity (Hanle-Zeeman regime) and orientation, both 
deterministic and micro-structured.
Working within the framework of a rigorous PRD theoretical approach, we have 
developed a numerical code which solves the full non-LTE radiative transfer 
problem for polarized radiation, in one-dimensional models of the solar 
atmosphere, accounting for the combined action of the Hanle and Zeeman effects, 
as well as for PRD phenomena.
After briefly discussing the relevant equations, we describe the
iterative method of solution of the problem and the numerical tools that we
have developed and implemented.
We finally present some illustrative applications to two resonance lines that form at 
different heights in the solar atmosphere, and provide a detailed physical 
interpretation of the calculated Stokes profiles. 
We find that in strong resonance lines sensitive to PRD effects 
the magneto-optical $\rho_V$ terms of the Stokes-vector transfer equation produce conspicuous  
$U/I$ wing signals along with a very interesting magnetic sensitivity in the wings of the linear polarization profiles. 
We also show that the weak-field approximation has
to be used with caution when PRD effects are considered.
\end{abstract}

\section{Introduction}
\label{sec_intro}
The most important physical observable for probing the thermal, dynamic, and 
magnetic properties of stellar atmospheres is the emerging radiation.
Aside from its intensity, the radiation is characterized by a given 
polarization state, which contains crucial information about the magnetic 
fields present in the atmosphere.
Although the magnetic field is known to play a key role in the atmosphere of the Sun and 
other stars, our empirical knowledge of its intensity and orientation is still 
largely unsatisfactory, and basically limited to the deepest layers (the 
photosphere).
This explains the importance of developing new techniques for magnetic field 
diagnostics, based on the accurate measurement and interpretation of the 
polarization properties of the radiation field.
Each line of the solar spectrum gives information on the physical properties of 
the solar atmosphere at a certain height range, depending on the opacity of the 
atmosphere at the frequency of the line in question. 
As examples, the Sr~{\sc i} line at 4607~{\AA} can be used to obtain 
information on the Sun's photosphere \citep[e.g.,][]{Trujillo+04},
while the lower chromosphere can be studied 
via the Sr~{\sc ii} line at 4078~{\AA} \citep[e.g.,][]{Bianda+98}.
Interpreting these Stokes profiles requires solving a radiative transfer 
problem out of local thermodynamic equilibrium (non-LTE), which becomes more 
complex if, besides the well-known Zeeman effect, we wish to model the impact 
of scattering polarization and its modification due to the presence of a 
magnetic field (Hanle effect). \newline
A solid theory for the generation and transfer of polarized radiation, based on 
a first-order perturbative expansion of the atom-radiation interaction within 
the framework of quantum electrodynamics, is today available 
\citep[e.g.,][hereafter LL04]{BLandiLandolfi2004}.
Within this theory, the scattering of a photon, which is intrinsically a 
second-order process, is described as a temporal succession of independent 
absorption and re-emission processes (Markov approximation).
This case, generally referred to as the limit of complete frequency 
redistribution (CRD) is strictly correct either when collisions are extremely 
efficient in relaxing any possible correlation between the frequencies of the 
incoming and outgoing photons, or when the pumping field is spectrally flat 
\citep[e.g.,][]{CasiniLandi07}.
Nonetheless, it can be shown that even when the above-mentioned conditions are 
not strictly verified, the limit of CRD represents in any case a suitable 
approximation for modeling the center of the spectral lines, where the Hanle 
effect takes place.
This theory, on the other hand, turns out to be unsuitable to model the wings 
of strong spectral lines, where coherent scattering and partial frequency 
redistribution (PRD) effects play a fundamental role.
Different theoretical approaches suitable to describe coherent scattering 
processes have been proposed during the last years.\footnote{ By 
``coherent scattering'' we mean here a scattering process in which the 
frequencies of the absorbed and emitted photons are either identical (if the 
initial and final states coincide), or satisfy the Raman scattering rule (if 
the initial and final states differ). In this sense, coherent scattering is 
strictly valid in the atomic reference frame, when the atom does not interact 
with any other particle (collisionless regime), and when the lower level can 
be assumed to be infinitely sharp (which is generally a good approximation 
when this is either the ground or a metastable level).}
One is based on the Kramers-Heisenberg scattering formula. This approach was 
initially proposed by \cite{Stenflo94}, and it has been recently extended to 
increasingly complex atomic models in the presence of arbitrary magnetic fields 
\citep[e.g.,][]{Sowmya14,Sowmya15}.
Another approach, also suitable to describe complex atomic models in the
presence of arbitrary magnetic fields, is based on the heuristic idea of 
metalevels \citep[see][]{Landi+97}.
A new quantum mechanical approach, capable of considering higher-order 
processes through a diagrammatic treatment of the atom-radiation interaction, 
has been recently proposed by \cite{Casini+14}.\newline
The coherency of scattering can be relaxed through two different physical 
mechanisms: the Doppler effect and collisional processes.
Doppler redistribution must always be considered when going from the atomic 
frame to the observer's one. Its inclusion in the above-mentioned approaches 
does not present particular difficulties from the theoretical point of view, 
although it leads to rather complex mathematical expressions.
On the contrary, the generalization of these approaches so to include 
collisional processes is not trivial, and it is still under investigation. A theoretical 
approach based on a perturbative expansion of the atom-radiation 
interaction, which includes collisional redistribution, has been proposed 
by \cite{Bommier97a,Bommier97b} for the case of a two-level atom.
This approach, which is based on the redistribution matrix formalism, is the 
starting point of our work.
We consider a two-level model atom with an unpolarized and infinitely sharp 
lower level. This atomic model is not only of academic interest, but is 
suitable to model various strong resonance lines of diagnost relevance, such as 
the Sr~{\sc i} line at 4607~{\AA}, or the Ca~{\sc i} line at 4227~{\AA}. 
Indeed, we observe that the lower levels of these lines, having total angular 
momenta $J=1/2$ and $J=0$, respectively, cannot be polarized (in particular, 
they cannot carry atomic alignment) by definition\footnote{Note that levels 
with $J=1/2$ can be polarized (they can carry atomic orientation) if the 
incident radiation is circularly polarized. In this work, we asume that 
collisional depolarization is always sufficiently strong so to destroy any 
atomic orientation that might be induced in the (long-lived) lower 
level of these resonance lines}. 
Moreover, these levels are the ground levels of the 
corresponding atomic species, so that the assumption that they are infinitely 
sharp is a very good approximation. 

In this work the solar atmosphere is modeled as one-dimensional, static, and 
plane-parallel. 
Though considering the atmosphere as dynamic and three-dimensional is a much 
more realistic treatment for the generation and transfer of polarized 
radiation, the approach presented here is a suitable first step, in which much 
faster calculations can be performed, yielding many insights into the physical 
mechanisms involved.

In Sect.~\ref{sec_form}, we present the starting equations, written in the 
atomic reference frame, with the quantization axis directed along the magnetic 
field, and we discuss their transformation into an arbitrary reference frame.  
Obtaining the emergent intensity and polarization requires finding the 
self-consistent solution of the statistical equilibrium (SE) equations for the 
atomic state, and of the radiative transfer (RT) equations.\footnote{ This 
investigation is carried out within the framework of the redistribution matrix 
formalism. 
We recall that the redistribution matrix is based on an analytical solution of 
the SE equations, which therefore do not explicitly appear in the problem,
when this formalism is applied.}
This is done through an iterative method that is described in 
Sect.~\ref{sec_iter}, together with the numerical tools that have been 
developed and implemented, considering the particular characteristics of the 
problem under investigation. 
When PRD phenomena are taken into account,
the emitted radiation at a given frequency does not depend only on the incoming 
radiation at that specific frequency (as would happen for coherent scattering), 
nor does it depend on a frequency-averaged radiation field 
(as in CRD).
In consequence, the iterative scheme has to consider that all frequencies are 
coupled to one another in the scattering process. 
The possiblity of having a magnetic field which is micro-structured, and its 
effect in this radiative transfer problem is discussed in 
Sect.~\ref{sec_micro}.
In Sect.~\ref{sec_res} we present some illustrative applications to some lines 
of diagnostic interest, based on the theory and numerical methods discussed in 
this paper.
\section{Formulation of the problem}
\label{sec_form}
In this work we consider a two-level atom with an unpolarized and 
infinitely-sharp lower level. 
The general RT equation that we need to solve in order to find 
the polarized radiation emerging from a stellar atmosphere can be written as
\begin{equation}
\frac{\mathrm{d}}{\mathrm{d}s}
\left( \begin{array}{c}
	I \\
        Q \\      
        U \\    
        V
\end{array} \right) = 
\left( \begin{array}{c}
        \varepsilon_I \\
        \varepsilon_Q \\
        \varepsilon_U \\
        \varepsilon_V 
\end{array} \right) - 
\left( \begin{array}{ c c c c }
	\eta_I & \eta_Q & \eta_U & \eta_V \\
	\eta_Q & \eta_I & \rho_V & -\rho_U \\
	\eta_U & -\rho_V & \eta_I & \rho_Q \\
	\eta_V & \rho_U & -\rho_Q & \eta_I \\
\end{array} \right) 
\left( \begin{array}{c}
	I \\
	Q \\
	U \\
	V 
\end{array} \right)
\label{RTE}                                       
\end{equation}
where $I$, $Q$, $U$, and $V$ are the four Stokes parameters, and $s$ is 
the spatial coordinate along the ray path.
 The quantities $\varepsilon_X$ ($X=I,Q,U$, and $V$) are the emission 
coefficients in the four Stokes parameters, the coefficients $\eta_X$ 
describe the differential absorption of the various polarization states 
(dichroism), while the coefficients $\rho_X$ describe couplings between 
different Stokes parameters (anomalous dispersion effects).
 As is well known, stimulated emission is negligible in the solar atmosphere, and
so will not be taken into account in this work.
The Stokes parameters and the RT coefficients are in general functions of the 
spatial point, and of the frequency ($\nu$) and propagation direction 
($\vec{\Omega}$) of the radiation beam under consideration.
The RT coefficients depend on the state of the atoms that, in non-LTE 
conditions, has to be calculated by solving the SE equations. 
When polarization phenomena are considered, it is necessary to provide a 
complete description of the atomic state, by specifying the population of the 
various magnetic sublevels as well as the quantum interference (or coherence) 
that may be present between pairs of them.
Whenever the magnetic sublevels are not evenly populated and/or quantum 
interference between pairs of them is present, the atomic level is said to 
be polarized.
In general, the four Stokes parameters are coupled to one another, and we solve
the transfer equation numerically, by applying a short-characteristics method 
known as DELOPAR (see \citealt{Trujillo03b}). 
\begin{figure}[htp]
\centering
\includegraphics[width=0.45\textwidth]{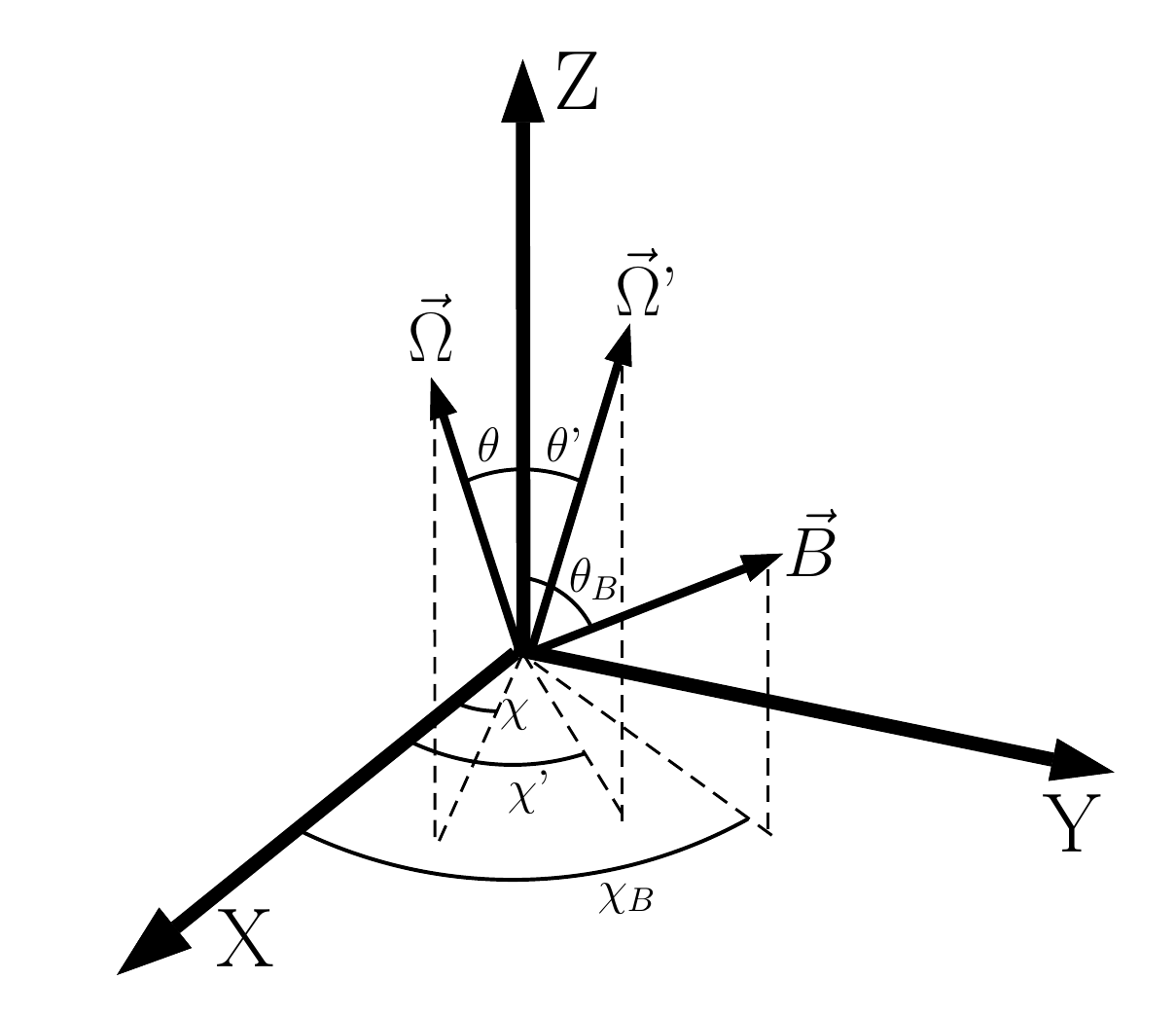}
\caption{Geometry of the problem. 
We take a right-handed Cartesian coordinate system with the Z-axis (quantization axis for the angular momentum)
along the local vertical, and the X-axis directed so that the line of sight towards the observer lies in the X-Z plane.
 In this reference system, the direction of the magnetic field $\vec{B}$ is specificied by its inclination with respect to the
  vertical ($\theta_B$) and its azimuth ($\chi_B$). Similarly, the direction of the incoming photon ($\vec{\Omega}'$) is specified by the 
  angles ($\theta'$,$\chi'$) and for the outgoing photon the direction ($\vec{\Omega}$) is specified by angles ($\theta$,$\chi$)}
\label{GeomFig}
\end{figure}

In general, the RT coefficients appearing in Eq.~\eqref{RTE} contain 
contributions due to both line and continuum processes. 
Hereafter, the line and continuum contributions will be 
distinguished through the labels $\ell$ and $c$, respectively.
Observing that dichroism and anomalous dispersion effects are 
generally negligible in the continuum spectrum, we have
\begin{equation}
	\eta^c_i(\nu) = \delta_{i 0} \, \eta^c_I(\nu) \, ,
\end{equation}
with $i=0,1,2$, and 3, standing for Stokes $I$, $Q$, $U$, and $V$, 
respectively, and
\begin{equation}
	\rho^c_i(\nu) = 0 \, , \quad (i=1,2,3) \, .
\label{ContCoef}
\end{equation}
It is worth noting that the continuum contribution to $\eta_I$ has a frequency 
dependence, but is independent of the direction of propagation of the 
radiation.
 Considering the contributions to the continuum due to both scattering 
processes, which we assume to be coherent, and thermal processes, the 
emission coefficient is given by (see LL04):
\begin{align}
	\varepsilon^c_i(\nu,\vec{\Omega}) = \sigma(\nu) \sum_{KQ} 
	{\mathcal T}^K_Q(i,\vec{\Omega}) \, (-1)^Q \, J^K_{-Q}(\nu) + 
	\varepsilon_{th}(\nu) \, \delta_{i 0} \; .
\end{align}
where $\sigma(\nu)$ is the continuum scattering cross section and 
$\varepsilon_{th}(\nu)$ is the thermal continuum emission coefficient. 
The quantity ${\mathcal T^K_Q(i,\vec{\Omega})}$ is the so-called
polarization tensor (see Sect.~5.11 of LL04), an irreducible 
spherical tensor of rank $K=0,1,2$ ($Q$ is an integer ranging from $-K$ to 
$K$).
The radiation field tensor, $J^K_Q(\nu)$, which provides a complete 
description of the symmetry properties of the radiation field, is 
defined by
\begin{equation}
	J^K_Q(\nu) = \oint \! \frac{\mathrm{d}\vec{\Omega}}{4\pi} \, 
	\sum_{j=0}^3 {\mathcal T}^K_Q(j,\vec{\Omega}) I_j(\nu,\vec{\Omega}) 
	\; ,
\label{radtens}
\end{equation}
where $I_j(\nu,\vec{\Omega})$ is the Stokes vector (i.e., a vector whose 
components are the four Stokes parameters). 

The line contributions to $\eta_i$ and $\rho_i$ are calculated according to 
LL04.
For the particular case of a two-level atom with an unpolarized lower level, 
in the atomic rest frame, and taking the quantization axis for the 
angular momentum along the direction of the magnetic field we have 
\begin{subequations}
\begin{align}
	\eta^\ell_i(\nu,\vec{\Omega}) & = k_L \sum_{K} 
	\Phi^{0 K}_0(J_\ell,J_u;\nu) {\mathcal T}^K_0(i,\vec{\Omega}) \, ,
\label{eta_ell} \\
	\rho^\ell_i(\nu,\vec{\Omega}) & = k_L \sum_{K} 
	\Psi^{0 K}_0(J_\ell,J_u;\nu) {\mathcal T}^K_0(i,\vec{\Omega}) \; , 
\label{Propag}
\end{align}
\end{subequations}
where $k_L$ is the frequency-integrated absorption coefficient, defined 
by
\begin{equation}
	k_L = \frac{h \nu}{4 \pi} {\mathcal N}_\ell B_{\ell u} \; ,
\label{kLcoef}
\end{equation}
where $h$ is the Planck constant, ${\mathcal N}_\ell$ is the population 
of the lower level, and $B_{\ell u}$ is the Einstein coefficient for 
absorption.  
 The quantities $\Phi^{K K'}_Q(J_\ell,J_u;\nu)$ and 
$\Psi^{K K'}_Q(J_\ell,J_u;\nu)$ are the so-called generalized profile 
and generalized dispersion profile, respectively, defined as:
\begin{align}
&\Phi^{K K'}_Q (J_\ell, J_u; \nu) = \sqrt{3 (2J_u +1) (2 K + 1) (2 K' +1)} \notag \\
& \times \sum_{M_u {M_u}' M_\ell q q'} (-1)^{1 + J_u - M_u + q'} 
\left(\begin{array}{c c c} 
J_u & J_\ell & 1 \\
-M_u & M_\ell & -q  \end{array} \right) 
 \left(\begin{array}{c c c} 
J_u & J_\ell & 1 \\
-{M_u}' & M_\ell & -q' \end{array} \right) \notag \\ 
& \times \left(\begin{array}{c c c}
J_u & J_u & K \\
{M_u}' & -M_u & -Q
 \end{array} \right) 
  \left(\begin{array}{c c c}
 1 & 1 & K' \\
 q & -q' & -Q \end{array} \right)  \frac{1}{2} \Bigl[\Phi(\nu_{M_u M_\ell} - \nu) + \Phi(\nu_{{M_u}' M_\ell} - \nu)^\ast \Bigr] \; ,
\label{GenProf}
\end{align}
and
\begin{align}
& \Psi^{K K'}_Q  (J_\ell, J_u; \nu) = \sqrt{3 (2J_u +1) (2 K + 1) (2 K' +1)} \notag \\
& \times \sum_{M_u {M_u}' M_\ell q q'} (-1)^{1 + J_u - M_u + q'} 
\left(\begin{array}{c c c} 
J_u & J_\ell & 1 \\
-M_u & M_\ell & -q  \end{array} \right)
\left(\begin{array}{c c c} 
J_u & J_\ell & 1 \\
-{M_u}' & M_\ell & -q' \end{array} \right) \notag \\
& \times  \left(\begin{array}{c c c}
J_u & J_u & K \\
{M_u}' & -M_u & -Q
 \end{array} \right)
\left(\begin{array}{c c c}
 1 & 1 & K' \\
 q & -q' & -Q \end{array} \right) \frac{(-\mathrm{i})}{2} \Bigl[\Phi(\nu_{M_u M_\ell} - \nu) 
 - \Phi(\nu_{{M_u}' M_\ell} - \nu)^\ast \Bigr] \; ,
  \label{DisProf}
\end{align}
where $J_u$ and $J_{\ell}$ are the total angular momenta of the upper and 
lower level, respectively, while $M_u$ and $M_{\ell}$ are the magnetic quantum 
numbers for the Zeeman sublevels of the upper and lower level, respectively. The
frequencies $\nu_{M_u M_\ell}$ are defined as
\begin{equation*}
\nu_{M_u M_\ell} = \frac{E(M_u) - E(M_\ell)}{h} \, ,
\end{equation*}
 where $E(M_u)$ and $E(M_\ell)$ are the energies of the magnetic sublevels $M_u$ and $M_\ell$, respectively. The indices can take the following values
\begin{align}
	K &= 0, 1,..., 2J_u \, , \notag \\
	K'&= 0, 1, 2 \, , \\
	Q &= 0, \pm 1, \pm 2 \, , \quad |Q| \le K \, , \quad |Q| \le K' 
	\notag \, .
\end{align}
The $\Phi$ profiles are defined by
\begin{equation*}
	\Phi(\nu_0 - \nu) = \phi(\nu_0 - \nu) + \mathrm{i} \psi(\nu_0-\nu) \; ,
\end{equation*}
where, in the atomic reference frame, $\phi(\nu_0-\nu)$ is the Lorentzian 
profile and $\psi(\nu_0-\nu)$ the associated dispersion profile.

Working within the framework of the redistribution matrix formalism, the line 
part of the emission coefficient is given by
\begin{align}
	& \varepsilon^\ell_i(\nu,\vec{\Omega}) = \, k_L \int_0^\infty \! 
	\mathrm{d}\nu' \oint \! \frac{\mathrm{d} \vec{\Omega}'}{4\pi} \, 
	\sum_{j=0}^3 \biggl[ 
	{\mathcal R}(\nu',\vec{\Omega}',\nu,\vec{\Omega};\vec{B})
	\biggr]_{ij}  I_j(\nu',\vec{\Omega}') 	\notag \\
	& + k_L \frac{\epsilon'}{1 + \epsilon'} 
	B_{T}(\nu_0) \sum_{K} {\mathcal T}^{K}_0(i,\vec{\Omega}) 
	\Phi^{0 K}_0(J_\ell,J_u;\nu) \; . 
\label{EmisLine}
\end{align}
The second term in the right-hand side of Eq.~\eqref{EmisLine}, generally 
referred to as the collisional (or thermal) term, describes the contribution 
to the emission coefficient due to atoms excited by isotropic collisions. This term 
depends on the parameter $\epsilon' = C_{u \ell}/A_{u \ell}$, with $C_{u \ell}$ 
the inelastic collisional de-excitation rate and $A_{u \ell}$ the Einstein 
coefficient for spontaneous emission. The quantity $B_{T}(\nu_0)$ is the 
Planck function in the Wien limit (consistently with our assumption of 
neglecting stimulated emission) at the line-center frequency $\nu_0$, 
 and at the temperature $T$. 
The radiative part of the emission coefficient (first term in the 
righthand side of Eq.~\ref{EmisLine}) contains the redistribution matrix
$\bigl[\mathcal{R}(\nu',\vec{\Omega}',\nu,\vec{\Omega};\vec{B})\bigr]_{ij}$, 
 where the primed quantities refer to the incoming radiation, while the 
unprimed ones refer to the outgoing radiation.
The redistribution matrix allows us to relate the emission coefficient 
directly to the Stokes parameters of the incoming radiation, a circumstance 
that is only possible when an analytical solution of the SE equations is 
available.
It can be shown that the most general form of the redistribution matrix 
is given by the linear combination of two terms, one describing purely 
coherent scattering (CS) in the atomic rest frame, and one describing 
scattering processes in the limit of CRD (following the terminology 
introduced by Hummer (1962) these
redistribution matrices are generally indicated with the symbols
${\mathcal R}_{\mbox{\sc \footnotesize II}}$ and 
${\mathcal R}_{\mbox{\sc \footnotesize III}}$, respectively):
\begin{equation}
  \biggl[ {\mathcal R}(\nu',\vec{\Omega}',\nu,\vec{\Omega};\vec{B})
	\biggr]_{ij}  = 
	\biggl[{\mathcal R}_{\mbox{\sc \footnotesize II}}(\nu',\vec{\Omega}',
	\nu,\vec{\Omega};\vec{B}) 
	\biggr]_{ij} + \biggl[ 
	{\mathcal R}_{\mbox{\sc \footnotesize III}}(\nu',\vec{\Omega}',
	\nu,\vec{\Omega};\vec{B}) 
	\biggr]_{ij} \; ,
\label{Redis}
\end{equation}

The details of the atom-radiation interaction, and therefore the relevant 
physics of partial frequency redistribution phenomena, are contained in the 
redistribution matrix.
In this work, we consider the redistribution matrix derived by 
\cite{Bommier97a,Bommier97b} for the case of a two-level atom with unpolarized 
and infinitely-sharp lower level, in the presence of arbitrary magnetic fields.
This redistribution matrix accounts for the various effects of elastic 
collisions, namely, level broadening, relaxation of atomic polarization, and 
frequency redistribution in the scattering processes.
We observe that the assumption of infinitely-sharp lower level is valid 
whenever the lifetime of the lower level is very large, and therefore is 
perfectly suitable for resonance lines, since their lower level is, 
by definition, the ground level.
In the atomic rest frame, and taking the quantization axis along the magnetic 
field, the ${\mathcal R}_{\mbox{\sc \footnotesize II}}$ and 
${\mathcal R}_{\mbox{\sc \footnotesize III}}$ redistribution matrices derived 
by \citet{Bommier97b}
have the following expressions:\footnote{The approximation 
\unexpanded{$\nu/\nu_0 \approx 1$} has been used, which is valid because, at 
frequencies significantly different from \unexpanded{$\nu_0$, 
$\varepsilon^\ell_i(\nu,\vec{\Omega})\rightarrow 0$}.}
\begin{align}
	& \biggl[{\mathcal R}_{\mbox{\sc \footnotesize II}}(\nu',\vec{\Omega}',
	\nu,\vec{\Omega}; \vec{B})\biggr]_{ij} = &\notag \\
	& \sum_{K' K'' Q} \,
	\sum_{\substack{M_u {M_u}' M_\ell {M_\ell}'\\ p p' p'' p'''}} 
	{\mathcal C}_{K' K'' Q M_u {M_u}' M_\ell {M_\ell}' p p' p'' p'''} 
	 \frac{\Gamma_R}{\Gamma_R + \Gamma_I + \Gamma_E + \mathrm{i} 
	\omega_L g_{J_u} Q} & \notag \\
	& \times (-1)^Q \, {\mathcal T}^{K''}_Q(i,\vec{\Omega}) \, 
	{\mathcal T}^{K'}_{-Q}(j,\vec{\Omega}') \, 
	\delta(\nu-\nu'-\nu_{M_\ell {M_\ell}'}) 
    \frac{1}{2} \Biggl[ \Phi(\nu_{{M_u}' M_\ell} - \nu') +
	\Phi^\ast(\nu_{{M_u} M_\ell} - \nu') \Biggr] \, ,
 \label{R2A}
\end{align}
\begin{align}
	& \biggl[{\mathcal R}_{\mbox{\sc \footnotesize III}}
	(\nu',\vec{\Omega}',\nu,\vec{\Omega}; \vec{B})\biggr]_{ij} = \notag \\
 	& \sum_{K K' K'' Q} \Biggl[ 
	\frac{\Gamma_R}{\Gamma_R + \Gamma_I + D^{(K)} + \mathrm{i} \omega_L 
	g_{u} Q}  - 
	\frac{\Gamma_R}{\Gamma_R + \Gamma_I + \Gamma_E + \mathrm{i} \omega_L 
	g_{u} Q}\Biggr] \notag\\
 	& \times (-1)^Q \, {\mathcal T}^{K''}_Q(i,\vec{\Omega}) \, 
	{\mathcal T}^{K'}_{-Q}(j,\vec{\Omega}') \,
     \Phi^{K K''}_Q(J_\ell,J_u;\nu) \,
	\Phi^{K K'}_Q(J_\ell,J_u;\nu') \, ,
\label{R3A}  
\end{align}
where $\Gamma_R$, $\Gamma_I$ and $\Gamma_E$ are the line broadening constants 
for radiative decays, collisional de-excitation and elastic collisions 
respectively:
\begin{equation*}
	\Gamma_R = A_{u \ell} \, , \quad 
	\Gamma_I = C_{u \ell} \, , \quad 
	\Gamma_E = Q_{\mbox{\footnotesize el}} \, ,
\end{equation*}
with $Q_{\mbox{\footnotesize el}}$ the elastic collision rate.
The rate $D^{(K)}$ is the $K-$multipole component of the depolarizing 
rate due to elastic collsions, $\omega_L$ is the angular Larmor frequency, and 
$g_{u}$ is the Land\'e factor of the upper level.
 The quantity
 ${\mathcal C}_{K' K'' Q M_u {M_u}' M_\ell {M_\ell}' p p' p'' p'''}$ is a real number which 
 depends on the indices and quantum numbers 
indicated as pedices.
Its explicit expression is given by \citep[see][]{Bommier97b}
\begin{align}
 & {\mathcal C }_{K' K'' Q M_u {M_u}' M_\ell {M_\ell}' p p' p'' p'''}  = 
 3 (2 J_u + 1) \sqrt{2K' + 1} \sqrt{2K'' + 1} (-1)^{2 J_u - M_\ell - {M_\ell}'} \notag \\
  &\times \left(\begin{array}{c c c} 
  J_u & J_\ell & 1 \\
  M_u & -M_\ell & -p \end{array} \right)
 \left(\begin{array}{c c c} 
  J_u & J_\ell & 1 \\
  {M_u}' & -{M_\ell} & -p' \end{array} \right)
 \left(\begin{array}{c c c}
 J_u & J_\ell & 1\\
 {M_u} & -{M_\ell}' & -p'' \end{array}\right) \notag \\
 & \times \left(\begin{array}{c c c}
 J_u & J_\ell & 1\\
 {M_u}' & -{M_\ell}' & -p'''
 \end{array} \right) 
 \left(\begin{array}{c c c} 
1 & 1 & K' \\
-p & p' & Q \end{array}\right)
\left(\begin{array}{c c c}
1 & 1 & K'' \\
-p'' & p''' & Q\end{array}\right) \; .
\label{Cfac}
\end{align}
\subsection{Expressions in an arbitrary reference frame} 
The previous expressions for the redistribution matrices are given in the 
magnetic reference frame, i.e., the frame in which the quantization axis 
is parallel to the direction of the magnetic field. 
However, one may want to express them in an arbitrary, fixed reference 
frame, by changing the direction of the quantization axis (e.g., taking it 
along the vertical direction for a plane-parallel atmosphere).
This transformation can be performed as described in Sect.~7.12 of LL04, 
taking into account the following rotation rule of the polarization 
tensor
\begin{equation}
	{\mathcal T}^K_{Q'}(i,\vec{\Omega})\biggl|_{\mbox{\footnotesize{new}}} 
	= \sum_{Q} 
	{\mathcal T}^K_Q(i,\vec{\Omega})\biggl|_{\mbox{\footnotesize{B}}} 
	{\mathcal D}^K_{Q Q'}(R_B) \; , 
\label{TauTensor}
\end{equation}
and the inverse relation
\begin{equation}
	{\mathcal T}^K_{Q'}(i,\vec{\Omega})\biggl|_{\mbox{\footnotesize{B}}} = 
 \sum_{Q} {\mathcal T}^K_Q(i,\vec{\Omega})\biggl|_{\mbox{\footnotesize{new}}} 
	{\mathcal D}^K_{Q' Q}(R_B)^\ast \; , 
\label{TauTensorInv}
\end{equation}
with $D^K_{Q Q'}(R_B)$ the rotation matrix, and $R_B$ the rotation
that brings the magnetic reference frame into the new reference frame.
For an arbitrary rotation $R = (\alpha,\beta,\gamma)$,  with 
$\alpha$, $\beta$ and $\gamma$ the Euler angles, the rotation matrix 
$D^K_{Q_1 Q_2}(R)$ is given by
\begin{equation}
	{\mathcal D}^K_{Q_1 Q_2}(R) = 
	\mathrm{exp} \biggl[ \mathrm{i}(\alpha Q_1 + \gamma Q_2) \biggr] 
	d^K_{Q_1 Q_2} (\beta) \; ,
\label{RotationMatr}
\end{equation}
where $d^K_{Q_1 Q_2}(\beta)$ is the so-called reduced rotation matrix, which 
is a real number that contains the information on the change in inclination of 
the system due to the rotation.
Referring to Fig.~\ref{GeomFig}, the rotation 
$R_B$ is defined by the Euler angles 
$R_B = (0,-\theta_B,-\chi_B).$\footnote{In full generality, there would be a 
third Euler angle $\alpha_B$. Nonetheless, it can be proven that the 
for the problem under consideration, the expressions of the rotation 
matrices
are independent of the choice of $\alpha_B$, which can thus be chosen to 
be zero.} 

Using Eqs.~\eqref{TauTensor} and~\eqref{TauTensorInv}, we can easily find 
the expressions of $\eta_i(\nu,\vec{\Omega})$, $\rho_i(\nu,\vec{\Omega})$,
 and $\varepsilon_i(\nu,\vec{\Omega})$ (for both line and continuum 
processes) in an arbitrary reference frame.
In particular, in the new reference frame, the redistribution matrices take
the form:
\begin{align}
	& \biggr[{\mathcal R}_{\mbox{\sc \footnotesize II}}(\nu',\vec{\Omega}',
	\nu,\vec{\Omega}; \vec{B}) \biggr]_{ij} =  \notag \\
	& \sum_{K' K'' Q Q' Q''} 
	\sum_{\substack{{M_u}' M_u M_\ell {M_\ell}'\\ p p' p'' p'''}} 
	{\mathcal C}_{K' K'' Q M_u {M_u}' M_\ell {M_\ell}' p p' p'' p''' } 
	 \frac{\Gamma_R}{\Gamma_R + \Gamma_I + \Gamma_E + \mathrm{i} 
	\omega_L g_{u} Q} \notag \\
	& \times (-1)^{Q'} \, {\mathcal T}^{K''}_{Q''}(i,\vec{\Omega}) \, 
	{\mathcal T}^{K'}_{-Q'}(j,\vec{\Omega}') {\mathcal D}^{K'}_{Q Q'}(R_B) \, 
	{\mathcal D}^{K''}_{Q Q''}(R_B) \notag \\
	& \times \delta(\nu-\nu'-\nu_{M_\ell {M_\ell}'}) 
	 \frac{1}{2}\Biggl[\Phi(\nu_{{M_u}' M_\ell} - \nu') + 
	\Phi(\nu_{{M_u} M_\ell} - \nu' )^\ast \Biggr] \; .
\label{R2B}
\end{align}
\begin{align}
	& \biggl[{\mathcal R}_{\mbox{\sc \footnotesize III}}(\nu',\vec{\Omega},
	\nu,\vec{\Omega}; \vec{B})\biggr]_{ij} = & \notag \\
	& \sum_{K K' K'' Q Q' Q''} \Biggl[ 
	\frac{\Gamma_R}{\Gamma_R + \Gamma_I + D^{(K)} + \mathrm{i} \omega_L 
	g_{u} Q} - \frac{\Gamma_R}{\Gamma_R + \Gamma_I + 
	\Gamma_E + \mathrm{i} \omega_L g_{u} Q} \Biggr] & \notag \\
	& \times (-1)^{Q'} \, {\mathcal T}^{K''}_{Q''}(i,\vec{\Omega}) \, 
	{\mathcal T}^{K'}_{-Q'}(j,\vec{\Omega}') 
	 {\mathcal D}^{K'}_{Q Q'}(R_B) \, 
	{\mathcal D}^{K''}_{Q Q''}(R_B)^\ast \notag \\ 
	& \times \Phi^{K K''}_Q(J_\ell,J_u;\nu) \, 
	\Phi^{K K'}_Q(J_\ell,J_u;\nu') \, .
 \label{R3B}  
\end{align}
Now it is useful to factorize the redistribution matrices as follows
\begin{equation}
   \biggl[ {\mathcal R}_{\mathrm{X}}(\nu',\vec{\Omega}',
	\nu,\vec{\Omega};\vec{B}) \biggr]_{i j} =
	 \sum_{K' K'' Q} \left[ R_{\rm X} \right]^{K' K''}_{Q}(\nu',\nu,B) \, 
	\left[ {\mathcal P}^{K' K''}_{Q}(\vec{\Omega}',\vec{\Omega},\hat{b}) 
	\right]_{ij} \, , 
\label{FactorRedis}
\end{equation}
with $\mathrm{X}= \mathrm{II}, \mathrm{III}$, and where the magnetic field has 
been indicated as $\vec{B} = B \hat{b}$. 
In this way, all the dependence of the redistribution matrix on the geometrical part 
of the problem (i.e., propagation directions of the incoming and outgoing radiation, and
the orientation of the magnetic field) is contained in the scattering phase matrix 
\begin{equation}
  \left[ {\mathcal P}^{K' K''}_{Q}(\vec{\Omega}',\vec{\Omega},\hat{b}) 
	\right]_{ij} = \sum_{Q' Q''} (-1)^{Q'} \, 
	{\mathcal T}^{K''}_{Q''}(i,\vec{\Omega}) \, 
	{\mathcal T}^{K'}_{-Q'}(j,\vec{\Omega}')
	 {\mathcal D}^{K'}_{Q Q'}(R_B) \, 
	{\mathcal D}^{K''}_{Q Q''}(R_B)^\ast \; .
\end{equation}
\subsection{Expressions in the observer's frame}
The expressions derived in the previous section are still valid in the 
atom's reference frame. We discuss now the transformation into the 
observer's frame, where Doppler redistribution has to be taken into 
account.
The Doppler effect only affects the frequency-dependent part of the 
redistribution matrix, the scattering phase matrix remaining unchanged. 
Assuming that the atoms have a Maxwellian distribution for both the 
thermal and microturbulent velocities, following \citet{BMihalas1978}, we find 
for $R_{\mbox{\sc \footnotesize II}}$ 
\begin{align}
	& \bigl[ R^{\mbox{\footnotesize{obs}}}_{\mbox{\sc \footnotesize II}}
	\bigr]^{K' K''}_{Q}(\nu',\nu,\Theta,B) =  \notag \\
	& \sum_{\substack{M_u {M_u}' M_\ell {M_\ell}'\\ p p' p'' p'''}} 
	{\mathcal C}_{K' K'' Q M_u {M_u}' M_\ell {M_\ell}' p p' p'' p'''} 
    \frac{\Gamma_R}{\Gamma_R + \Gamma_I + \Gamma_E + \mathrm{i} 
	\omega_L g_{J_u} Q} \notag \\
	& \times \frac{1}{\pi \Delta{\nu_D}^2} \frac{1}{\sin\Theta} 
	\mathrm{exp} \Biggl[ - \biggl( \frac{\nu'-\nu+\nu_{M_\ell {M_\ell}'}}
	{2 \Delta\nu_D \sin(\Theta/2)} \biggr)^2 \Biggr] \notag \\
	& \times \frac{1}{2} \Biggl[ 
	W \biggl( \frac{a}{\cos(\Theta/2)},\frac{x_{{M_u}' M_\ell} + 
	{x'}_{{M_u}' {M_\ell}'}}{2 \cos(\Theta/2)} \biggr)  + W \biggl( \frac{a}{\cos(\Theta/2)},\frac{x_{M_u M_\ell} +  
	{x'}_{M_u {M_\ell}'}}{2\cos(\Theta/2)} \biggr)^\ast \Biggr] \, ,
\label{R4B}
\end{align}
 where $\Theta$ is the scattering angle (i.e., the angle between the 
directions of the incoming and outgoing photons), and where the function 
$W$ is defined by:
\begin{equation}
	W(a,x) = H(a,x) + \mathrm{i} \, L(a,x) \; ,
\label{Wfunction}
\end{equation}
with $H$ the Voigt function, and $L$ the Faraday-Voigt dispersion profile.
The quantity $a = \Gamma / 4 \pi \Delta \nu_D$ is the damping parameter,
with $\Gamma = \Gamma_R + \Gamma_I + \Gamma_E$. The reduced frequencies are defined as:
\begin{equation}
	x_{M_u M_\ell} = \frac{\nu_{M_u M_\ell} - \nu}{\Delta \nu_D} \, , \; 
	{x'}_{M_u M_\ell} = \frac{\nu_{M_u M_\ell} - \nu'}{\Delta \nu_D} \, .
\label{RedFreq}
\end{equation}
Because of the presence of the angle $\Theta$ in the quantity
$\bigl[ R^{\mbox{\footnotesize obs}}_{\mbox{\sc \footnotesize II}} 
\bigr]^{K' K''}_{Q}(\nu',\nu,\Theta,B)$,  
the angular and frequency dependencies cannot be factorized (as could be done 
in the atomic rest frame), which makes the problem significantly 
more complicated from the numerical point of view. 
In order to simplify it, we 
follow \citet{ReesSaliba82}, and we consider the expression of $\bigl[ 
R^{\mbox{\footnotesize obs}}_{\mbox{\sc \footnotesize II}} 
\bigr]^{K' K''}_{Q}$ averaged over the scattering angle
 (angle-averaged approximation). Detailed information on the
 range of validity of this approximation can be found in
 \citet{Faurobert87,Faurobert88} in the absence of magnetic field. 
 For a discussion of the validity of this approximation in the presence
 of a weak magnetic field see \citet{Sampoorna+08} and \citet{Sampoorna11}.
 Using such approximation, the frequency-dependent part of the redistribution function becomes:
\begin{align}
& \bigl[R^{\mbox{\footnotesize obs}}_{\mbox{\sc \footnotesize II-AA}} \bigr]^{K' K''}_{Q}(\nu',\nu;B) = \notag \\
& \sum_{\substack{M_u {M_u}' M_\ell {M_\ell}'\\ p p' p'' p'''}} {\mathcal C}_{K' K'' Q M_u {M_u}' M_\ell {M_\ell}' p p' p'' p'''} 
\frac{\Gamma_R}{\Gamma_R + \Gamma_I + \Gamma_E + \mathrm{i} \omega_L g_{J_u} Q} \notag \\
& \times \frac{1}{2\pi \Delta{\nu_D}^2}\int_{0}^{\pi}\!\mathrm{d}\Theta\,
\mathrm{exp} \Biggl[- \biggl(\frac{\nu'-\nu+\nu_{M_\ell {M_\ell}'}}{2 \Delta\nu_D \sin(\Theta/2)}\biggr)^2\Biggr] \notag \\
& \times \frac{1}{2}\Biggl[ W\biggl(\frac{a}{\cos(\Theta/2)},\frac{x_{{M_u}' M_\ell} + {x'}_{{M_u}' {M_\ell}'}}{2 \cos(\Theta/2)}\biggr)
 + W\biggl(\frac{a}{\cos(\Theta/2)},\frac{x_{M_u M_\ell} + {x'}_{M_u {M_\ell}'}}{2\cos(\Theta/2)}\biggr)^\ast \Biggr] \; .
\label{R5B}
\end{align}
The integration over the scattering angle $\Theta$ is performed 
numerically, using a Gauss-Legendre quadrature rule.\footnote{ We observe 
that the presence of an imaginary part in the redistribution function does not 
allow a trivial generalization to the present case of approximate methods for 
the evaluation of this integral, such as that proposed by 
\citet{Gouttebroze86}}
The dependence on this angle is contained in the exponential and $W$ functions,
both of which become steeper as their arguments approach zero.
For this reason, the number of quadrature points has been chosen depending
on the considered frequencies of the incoming and outgoing photons.
A particularly high number of points (of the order of 100) has to be
considered when the frequencies of the incoming and outgoing photons are such
that the argument of the exponential is zero.
We have checked that the numerical relative error in the evaluation of this
integral remains always below $10^{-6}$.
Following the same approach for the $R_{\mbox{\sc \footnotesize III}}$ 
redistribution matrix leads to a rather complicated expression. 
For simplicity, we take the approximation that CRD occurs in the observer's 
frame, for which we simply convolute the generalized profiles in equation 
\eqref{R3B} with a Gaussian function in order to account for the thermal and 
microturbulent velocity distribution. Thus, we substitute the Lorentzian and 
associated dispersion profiles appearing in Eq.~\eqref{GenProf} by the 
Voigt and Faraday-Voigt functions, respectively.
The ensuing expression of $[ R_{\mbox{\sc \footnotesize III}} 
]^{K' K''}_{Q}$ will be indicated as 
$[ R^{\mbox{\footnotesize obs}}_{\mbox{\sc \footnotesize III-CRD}} 
]^{K' K''}_{Q}$.
The same susbtitutions in the generalized profiles are used in the $\eta_i$ 
and  $\rho_i$ RT coefficients and in the thermal part of the line 
emission coefficient when transforming them into the observer's reference 
system. 
\section{Iterative method}
\label{sec_iter}
In full generality, the solution of Eq.~\eqref{RTE} for a discrete spatial 
grid of $N_P$ points, for any given frequency $\nu$ and propagation 
direction $\vec{\Omega}$, after introducing the optical depth scale
$\mathrm{d}\tau = -\eta_I(\nu,\vec{\Omega}) \mathrm{d}s$, can be expressed as:
\begin{equation}
	I_i(\nu,\vec{\Omega};n) = \sum_{j=0}^3 \sum_{m=1}^{N_P} 
	\Lambda_{\nu,\vec{\Omega}}(n,m)_{ij} \, S_j(\nu,\vec{\Omega};m) + 
	T_i(\nu,\vec{\Omega};n) \; ,
\label{Grideq}
\end{equation}
where, with the letters $n$ and $m$, we have explicitly indicated the 
dependence of the various quantities on the spatial grid points.
The quantity $S_j(\nu,\vec{\Omega};m)$ is the so-called source function 
at spatial point $m$, defined as:
\begin{equation}
	S_i(\nu,\vec{\Omega};m) = \frac{\varepsilon_i(\nu,\vec{\Omega};m)}
	{\eta_I(\nu,\vec{\Omega};m)} \; .
\label{Source}
\end{equation}
Distinguishing between the line and continuum processes, the source function 
can be further written as:
\begin{align}
	S_i(\nu,\vec{\Omega};m) = & \, r(\nu,\vec{\Omega};m) 
	S^\ell_i(\nu,\vec{\Omega};m) \notag\\
	& + \bigl(1 - r(\nu,\vec{\Omega};m) \bigr) S^c_i(\nu;m) \; ,
\end{align}
where
\begin{align}
	S^\ell_i(\nu,\vec{\Omega};m) & = 
	\frac{\varepsilon^\ell_i(\nu,\vec{\Omega};m)}{
	\eta^\ell_I(\nu,\vec{\Omega};m)} \, , 
\label{SourceLine}
	\, \\
	S^c_i(\nu,\vec{\Omega};m) & = 
	\frac{\varepsilon^c_i(\nu,\vec{\Omega};m)}{\eta^c_I(\nu;m)} \, , 
\end{align}
and 
\begin{equation}
	r(\nu,\vec{\Omega};m) = \frac{\eta^\ell_I(\nu,\vec{\Omega};m)}
	{\eta^\ell_I(\nu,\vec{\Omega};m) + \eta^c_I(\nu;m)} \; .
\end{equation}
The quantity $T_i(\nu,\vec{\Omega};n)$ is the radiation transmitted from 
the boundaries to point $n$. For given values of $\nu$, $\vec{\Omega}$, $n$ and $m$,
$\Lambda_{\nu,\vec{\Omega}}(n,m)_{ij}$ is a formal $4 \times 4$ operator which
depends on the propagation matrix appearing in  Eq.~\eqref{RTE}.
Numerically, $\Lambda_{\nu,\vec{\Omega}}(n,m)_{ij}$ represents the contribution 
to $I_i(\nu,\vec{\Omega})$ at point $n$ due to a source function 
$S_j(\nu,\vec{\Omega})$ which is zero everywhere except at point $m$, where it 
has a value of $1$. 
Therefore, aside from the radiation transmitted from the boundaries, all 
information on the generation and transfer 
of radiation in the atmosphere is contained in the 
$\Lambda_{\nu,\vec{\Omega}}(n,m)_{ij}$ operator elements. 
In order to calculate these operator elements from the source function and 
propagation matrix, we have applied the DELOPAR formal solver 
(see Sect.~\ref{sec_form}).
When the angle-averaged approximation for $\mathcal{R}_{\rm II}$ and the 
assumption of CRD in the observer's frame for $\mathcal{R}_{\rm III}$ are 
considered, the line emission coefficient can be expanded as  

\begin{equation}
	\varepsilon^\ell_i(\nu,\vec{\Omega}) = \sum_{K'' Q''} 
	{\mathcal T}^{K''}_{Q''}(i,\vec{\Omega}) \, 
	{\mathcal E}^{K''}_{Q''}(\nu)^\ell \, .
\label{EmisExpans}
\end{equation}
A similar expansion cannot be written for the line source function since 
in the presence of magnetic fields of arbitrary intensity, the line part of 
the absorption coefficient, $\eta_I^{\ell}(\nu,\vec{\Omega})$, which appears in 
the denominator of the line source function (see Eq.~\ref{SourceLine}), is 
also given by a linear combination of terms depending on the propagation 
direction $\vec{\Omega}$ (see Eq.~\ref{eta_ell}).
We thus write the line source function at a given frequency, direction and 
spatial point in the atmosphere as
\begin{equation}
	S^\ell_i(\nu,\vec{\Omega}) = 
	\frac{\sum_{K'' Q''} {\mathcal T}^{K''}_{Q''}(i,\vec{\Omega}) 
	{\mathcal E}^{K''}_{Q''}(\nu)^\ell}{\eta^\ell_I(\nu,\vec{\Omega})} \, . \\
\end{equation}
Recalling the equations derived in the previous Section, it can be seen 
that the components ${\mathcal E}^{K''}_{Q''}(\nu)^{\ell}$ are given by
\begin{equation}
	{\mathcal E}^{K''}_{Q''}(\nu)^\ell =  {\mathcal J}^{K''}_{Q''}(\nu) 
	+ k_L \frac{\epsilon'}{1 + \epsilon'} B_{T}(\nu_0) 
  	\Phi^{0 K''}_0(J_\ell,J_u;\nu) \, 
	{\mathcal D}^{K''}_{0 Q''}(R_B)^\ast \; , 
\label{SourLine}
\end{equation}
where 
${\mathcal J}^K_Q(\nu)$ is defined as
\begin{equation}
	{\mathcal J}^{K''}_{Q''}(\nu) = k_L \sum_{K' Q Q'} (-1)^{Q'} \, 
	{\mathcal D}^{K'}_{Q Q'}(R_B) \, \mathcal{D}^{K''}_{Q Q''}(R_B)^\ast 
    \int_0^\infty \! \mathrm{d}\nu' \, J^{K'}_{-Q'}(\nu') \, 
	R^{K' K''}_{Q}(\nu',\nu;B) \, ,
\label{Jquant}
\end{equation}
with
\begin{equation}
	R^{K' K''}_{Q} = 
	\bigl[ R^{\mbox{\footnotesize obs}}_{\mbox{\sc \footnotesize II-AA}} 
	\bigr]^{K' K''}_{Q} +
	\bigl[ R^{\mbox{\footnotesize obs}}_{\mbox{\sc \footnotesize III-CRD}} 
	\bigr]^{K' K''}_{Q} \, .
\end{equation}
Recalling the expression for the radiation field tensor in Eq.~\eqref{radtens}, 
and writing the Stokes vector in terms of the source functions as in 
Eq.~\eqref{Grideq}, we can rewrite the ${\mathcal J}^K_Q$ at a given spatial 
point as:
\begin{align}
	& {\mathcal J}^{K''}_{Q''}(\nu;n) = \, k_L \sum_{K' Q Q'} 
	\int_0^\infty \! \mathrm{d}\nu' R^{K' K''}_{Q}(\nu',\nu;n) 
	 {\mathcal D}^{K'}_{Q Q'}(R_B) 
	\mathcal{D}^{K''}_{Q Q''}(R_B)^\ast \, (-1)^{Q'} 
	\oint \frac{\mathrm{d}\Omega'}{4\pi} \sum_{i=0}^3 
	{\mathcal T}^{K'}_{-Q'}(i,\vec{\Omega}') \notag \\
	& \times \biggl\{ \sum_{m = 1}^{N_p} \sum_{j=0}^3 
	\Lambda_{\nu',\vec{\Omega}'}(n,m)_{ij} \Biggl[ 
	\frac{\sum_{K_P Q_P} {\mathcal T}^{K_P}_{Q_P}(j,\vec{\Omega'})
	{\mathcal E}^{K_P}_{Q_P}(\nu')^\ell}{\eta^\ell_I(\nu',\vec{\Omega'})} 
	\Biggr] + T_i(\nu',\vec{\Omega}';n) \biggr\}\, ,
\label{Jqexp}
\end{align}

In this equation, and for the remainder of this section, the explicit 
dependence of $R^{K' K''}_{Q}(\nu',\nu;n)$ on the magnetic field strength is 
no longer indicated, 
as this is implicitly included in the label for the spatial point $n$. 

From the previous equations it can easily be seen how the RT problem can 
be solved iteratively. We start with an estimate of the radiation field tensor 
$J^K_Q$ at each frequency and spatial point in the atmosphere.
From this estimate, we calculate the 
emission coefficient $\varepsilon_i(\nu,\vec{\Omega})$ by means of
Eq.~\eqref{EmisExpans}. From the emission coefficient, we can get new 
values of $J^K_Q$ via a formal solution of the RT equations. 
This iterative scheme is known as the lambda iteration method which, while 
simple, has a very slow convergence rate in optically thick media. 
  
For this reason, for the line part of the source function we apply the Jacobi 
iterative method, following Sect.~3 of \citet{TrujilloManso99}. 
In order to simplify the notation, for the rest of this section we will 
omit the apex ``$\ell$'' on the quantity ${\mathcal E}^{K''}_{Q''}$, 
being implicit that it refers to the line contribution.
When applied to our problem, the Jacobi method basically consists in the 
following procedure.
At any grid point $n$, the tensor ${\mathcal J}^K_Q(\nu,n)$ is calculated 
through a formal solution of the RT equations, by using the values of 
${\mathcal E}^{K}_Q$ obtained at the end of the previous iteration, herafter 
$[{\mathcal E}^{K}_Q]^{\mbox{\tiny old}}$, at all grid points, except 
at point $n$ where the new values, $[{\mathcal E}^{K}_Q]^{\mbox{\tiny new}}$,
are implicitly used.
 Within the formalism previously introduced, this reads:
\begin{equation}
 {\mathcal J}^{K''}_{Q''}(\nu;n) = 
	\left[ {\mathcal J}^{K''}_{Q''}(\nu;n) \right]^{\mbox{\tiny old}} 
   + \int_0^\infty\! \mathrm{d}\nu' \Biggl\{ \sum_{K_P Q_P} 
	\Lambda_{K'' Q'', K_P Q_P}(\nu',\nu;n) \, 
	\Delta{\mathcal E}^{K_P}_{Q_P}(\nu';n) \Biggr\} \; ,
\label{Jquancomp}
\end{equation}
where $\Delta {\mathcal E}^K_Q(\nu;n) = 
[{\mathcal E}^K_Q(\nu,\vec{\Omega};n)]^{\mbox{\tiny new}} - 
[{\mathcal E}^K_Q(\nu,\vec{\Omega};n)]^{\mbox{\tiny old}}$, and where 
$[{\mathcal J}^K_Q(\nu,\vec{\Omega};n)]^{\mbox{\tiny old}}$ has been calculated 
according to Eq.~\eqref{Jquant}, considering the radiation field tensor 
$[J^K_Q]^{\mbox{\tiny old}}$ that is obtained from a formal solution of the RT 
equations, using $[{\mathcal E}^{K}_Q]^{\mbox{\tiny old}}$ at all spatial grid 
points, as in the Lambda iteration method.
 Finally, we have defined the operator:
\begin{align}
	& \Lambda_{K'' Q'', K_P Q_P}(\nu,\nu';n) = k_L \sum_{K' Q Q'} 
	(-1)^{Q'} R^{K' K''}_{Q}(\nu',\nu;n) {\mathcal D}^{K'}_{Q Q'}(R_B)
	 {\mathcal D}^{K''}_{Q Q''}(R_B)^\ast  \notag \\
 	& \times \oint\!\frac{\mathrm{d} \vec{\Omega}'}{4 \pi} 
	\frac{r(\nu',\vec{\Omega}';n)}{\eta_I(\nu',\vec{\Omega'};n)^\ell}
	\sum_{i,j = 0}^3 
	\Lambda_{\nu',\vec{\Omega}'}(n,n)_{ij} \,
	{\mathcal T}^{K'}_{-Q'}(i,\vec{\Omega}') \,
	{\mathcal T}^{K_P}_{Q_P}(j,\vec{\Omega}') \, ,
\label{LambdaOp} 
\end{align}
 The correction $\Delta {\mathcal E}^{K''}_{Q''}(\nu;n)$ is obtained by 
substituting Eq.~\eqref{Jquancomp} into Eq.~\eqref{SourLine}.

Now, if the magnetic field is weak enough so that $I \gg Q,U,V$, then 
the contribution from ${\mathcal E}^{0}_0$ will dominate over the others,
and it will be sufficient to consider the 
$\Lambda_{00,00}$ operator only
\begin{align}
	& \Delta {\mathcal E}^{0}_0(\nu;n) = \int\!\mathrm{d}\nu' \, 
	\Lambda_{00,00}(\nu,\nu';n) \Delta {\mathcal E}^{0}_0(\nu';n) + 
	{\mathcal J}^0_0(\nu;n)^{\mbox{\tiny old}} \notag \\
	& \;+ k_L \frac{\epsilon'}{1 + \epsilon'} B_{T}(\nu_0) 
	\Phi^{0 0}_0(J_\ell,J_u,\nu;n) - 
	{\mathcal E}^{0}_0(\nu;n)^{\mbox{\tiny old}} \, .
\label{ConvScheme1}
\end{align}
In this way, the Jacobi method is only applied for calculating the 
correction $\Delta {\mathcal E}^0_0(\nu;n)$, while for the rest of 
$\Delta {\mathcal E}^{K''}_{Q''}(\nu;n)$ lambda iteration is used.
However, when the magnetic field gets larger and the resulting polarization 
fraction starts to be more significant, this approximation becomes increasingly 
inaccurate and the convergence rate begins to deteriorate, even producing 
instabilities. 
The first improvement that can be considered is the following: we keep 
applying the Jacobi method only for calculating the correction to 
$\mathcal{E}^0_0$, but we take into account the effect that polarization 
has on it. Recalling Eq.~\eqref{Jqexp}, this requires considering
the contribution from the various ${\mathcal E}^{K_P}_{Q_P}$ with $K_P \neq 0$, 
$Q_P \neq 0$ or, equivalently, to consider all 
$\Lambda_{00, K_P Q_P}(\nu',\nu)$ in Eq.~\eqref{Jquancomp}:
\begin{align}
	& \Delta {\mathcal E}^{0}_0(\nu;n) = \int\!\mathrm{d}\nu' \, 
	\sum_{K_P Q_P} \Lambda_{0 0, K_P Q_P}(\nu,\nu';n)
	 \Delta {\mathcal E}^{K_P}_{Q_P}(\nu';n) + 
	{\mathcal J}^0_0(\nu;n)^{\mbox{\tiny old}} - 
	{\mathcal E}^{0}_0(\nu;n)^{\mbox{\tiny old}}  \notag \\
	& + k_L \frac{\epsilon'}{1 + \epsilon'} B_{T}(\nu_0)
	 \Phi^{0 0}_0(J_\ell,J_u,\nu;n) 
\label{ConvScheme2}
\end{align}
Note that this approximation is only applicable in the cases where, though 
contributions from $Q,U,V$ cannot be neglected, 
$\bigl|{\mathcal E}^{0}_0\bigl|$ is still substantially larger than other 
$\bigl|{\mathcal E}^{K}_{Q}\bigl|$. So, despite the fact that the 
${\mathcal E}^{K}_{Q}$ have a contribution from all Stokes parameters, 
their convergence can still be driven only by the change in 
${\mathcal E}^{0}_0$. 
However, for larger magnetic fields, for which 
$\bigl|{\mathcal E}^{K}_{Q}\bigl|$, with $K,Q \ne 0$, become comparable in 
magnitude to $\bigl|{\mathcal E}^{0}_0\bigl|$, the Jacobi iterative 
scheme needs to be applied to all components, and not just to
$\mathcal{E}^0_0$.
This, on the other hand, produces a complex coupling of the various 
multipolar components, as well as of the various frequencies, so that the 
calculation of the corrections $\Delta {\mathcal E}^{K''}_{Q''}(\nu;n)$ implies 
the solution of a huge system of equations.
This is in general a formidable numerical problem, which can no longer be 
solved in a reasonable amount of time without resorting to suitable
computational techniques.
\begin{figure*}[htp]
\centering
 \includegraphics[width=\textwidth]{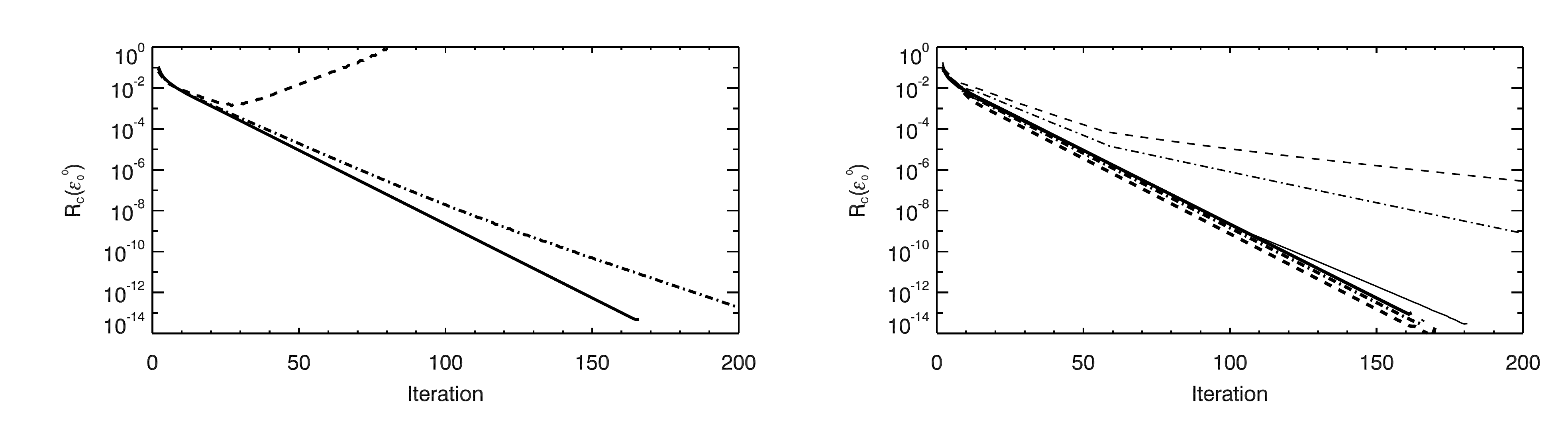}
 \caption{Convergence rates for ${\mathcal E}^{0}_0$ for various magnetic field strengths, for a deterministic magnetic field of inclination $\theta_B = \pi/2$ and $\chi_B = \pi/2$. Left: Convergence rate calculated using only the $\Lambda_{0 0,0 0}$ operator.
Right: Convergence rate calculated considering $\Lambda_{0 0, K_P Q_P}$. Results are plotted for field strengths of 0 G (thick solid line), 500 G (thick dash-dotted line), 800 G (thick dashed line), 1500 G (thin solid line), 2000 G (thin dash-dotted line), and 2500 G (thin dashed line).}
\label{ConvergRate}
\end{figure*} 
Whether one must consider all $\Lambda_{0 0, K_P Q_P}$ operators or it is sufficient to 
only consider $\Lambda_{00, 00}$, we are faced with a system of equations which 
we solve numerically in analogy to what is described in appendices D and E of 
\citet{BelluzziTrujillo14}. 
The systems for ${\mathcal E}^{0}_0$ appearing in Eq.~\eqref{ConvScheme1} or 
Eq.~\eqref{ConvScheme2} can be written, for every height point, in a more 
compact form as:
\begin{equation}
 \hat{M} \vec{\Delta {\mathcal E}^{0}_0} = \vec{\mathcal{C}} \; ,
\label{FBF}
\end{equation}
where $\hat{M}$ is a $N_F \times N_F$ matrix, with $N_F$ the number of 
points in the frequency grid, while $\vec{\Delta {\mathcal E}^{0}_0}$ and 
$\vec{\mathcal{C}}$ are vectors over the same grid. 
Notice that in both \eqref{ConvScheme1} and \eqref{ConvScheme2}, the dependence 
on $\Delta {\mathcal E}^{0}_0$ is the same and only $\mathcal{C}(\nu)$ changes. 
When considering only $\Lambda_{00,00}$, 
\begin{equation}
	\mathcal{C}(\nu;n) = {\mathcal J}^{0}_0(\nu;n)^{\mbox{\tiny old}} 
	- {\mathcal E}^{0}_0(\nu;n)^{\mbox{\tiny old}}
 	+ k_L \frac{\epsilon'}{1 + \epsilon'} B_{T}(\nu_0) 
	\Phi^{0 0}_0(J_\ell,J_u,\nu;n) \, ,
\label{FBFRho}
\end{equation}
while when we take into account contributions from all 
$\Lambda_{0 0,K_P Q_P}(\nu',\nu;n)$, $\mathcal{C}(\nu;n)$ is calculated with 
additional terms related to $\Delta {\mathcal E}^{K''}_{Q''}$, and therefore 
requires much more time per iteration. Information on other methods for
the transfer of spectral line polarization accounting for PRD effects 
can be found in \citet{Nagendra+02} and \citet{Sampoorna+08}\newline

We have analyzed the convergence rate of our method, and its dependence on the 
magnetic field strength, for a deterministic field with 
$\theta_B = \pi/2$ and $\chi_B = \pi/2$, applied to the Sr~{\sc ii} line at 
$4078$~\AA\ in 
the atmospheric model C of \citet{Fontenla+93}. 
The convergence rate is quantified through the maximum relative change of 
the ${\mathcal E}^{K}_{Q}(\nu;n)$ quantities:
\begin{equation}
	R_c({\mathcal E}^{K}_{Q}) = \mbox{max} \biggl( 
	\frac{ \bigl| {\mathcal E}^K_Q(\nu;n)^{\mbox{\tiny new}} - 
	{\mathcal E}^K_Q(\nu;n)^{ \mbox{\tiny old}} \bigr| }
	{\bigl| {\mathcal E}^K_Q(\nu;n)^{\mbox{\tiny new}} \bigr|} 
	\biggr) \, ,
\label{MaxRelChange}
\end{equation}
with the maximum evaluated over all frequencies and atmospheric heights.
The variation of $R_c({\mathcal E}^{0}_0)$ with the iteration 
number is shown in Fig.~\ref{ConvergRate} for the Sr {\sc ii} line at $4078$ \AA\ , where we compare the 
convergence rates found using only the $\Lambda_{00,00}$ operator (left panel) 
and using all $\Lambda_{00,K_P Q_P}$ (right panel).
Up to around a few hundred gauss both methods perform similarly well, but as 
the field strength is increased up to around 500~G, the convergence rate 
as calculated using $\Lambda_{00,00}$ only begins to deteriorate, and at 
800~G it produces instabilities.
However, when using $\Lambda_{0 0, K_P Q_P}$ the convergence rate does not begin 
to deteriorate until around 1500~G. Then, instabilities are also encountered in 
this case, just above 2500~G, and in order to proceed to larger field 
strengths, it would be necessary to perform a computationally expensive 
calculation considering all $\Lambda_{K'' Q'', K_P Q_P}$ operators.
In these calculations, we have considered model C of Fontenla et al. (1993), which is composed of 70 height points. We have used a frequency grid of 183
points. The grid is finer in the line core, where the points are equally
spaced, and coarser in the wings, where the separation among the points
increases logarithmically.
Some extra points have been added  
where the polarization profiles show abrupt changes (e.g., the $Q/I$ wing peaks).
As far as the integral over the propagation directions of the incoming radiation
is considered, we have applied a Gauss-Legendre quadrature over the
inclinations, considering 18 angles, 9 in the $[0,\frac{\pi}{2}]$ 
interval and 9 in the $[\frac{\pi}{2},\pi]$ interval. 
For the azimuthal integration we used the trapezoidal method with 8 angles.
\section{Micro-structured magnetic field}
\label{sec_micro} 
We consider also a unimodal micro-structured 
magnetic field  
i.e., a magnetic field of a given strength and an orientation that changes over 
scales below the line photon's mean free path. 
In this case, the RT coefficients  
appearing in Eq~\eqref{RTE} must be suitably averaged over 
the field directions:
\begin{equation}
 \frac{\mathrm{d}}{\mathrm{d}s}\left(\begin{array}{c}
                                                     I \\
                                                     Q \\      
                                                     U \\    
                                                     V
                                                   \end{array}\right)
 = \left(\begin{array}{c}
       \mean{\varepsilon_I} \\
       \mean{\varepsilon_Q} \\
       \mean{\varepsilon_U} \\
       \mean{\varepsilon_V} 
       \end{array} \right)
       - 
\left(\begin{array}{ c c c c }
  \mean{\eta_I} & \mean{\eta_Q} & \mean{\eta_U} & \mean{\eta_V} \\
  \mean{\eta_Q} & \mean{\eta_I} & \mean{\rho_V} & -\mean{\rho_U} \\
  \mean{\eta_U} & -\mean{\rho_V} & \mean{\eta_I} & \mean{\rho_Q} \\
  \mean{\eta_V} & \mean{\rho_U} & -\mean{\rho_Q} & \mean{\eta_I} \\
\end{array}\right) \left( \begin{array}{c}
 I \\
 Q \\
 U \\
 V
\end{array}\right) \; ,
\end{equation}
 where the symbol $\mean{...}$ indicates the above-mentioned average.
We shall now consider two cases: when the micro-structured field 
 is isotropic, and when its inclination 
is fixed but its azimuth changes over scales smaller than the line's photon 
mean free path.

\noindent
{\it a) Micro-structured isotropic field}:\\
In this case we need to average the RT coefficients 
over all inclinations $\theta_B$ and azimuthal directions $\chi_B$. 
Observing that the dependence of the RT coefficients on the magnetic 
field orientation is fully contained in the rotation matrices, the
problem reduces to the evaluation of the following integrals:
\begin{subequations}
\label{Miciso}
\begin{align}
	& \frac{1}{4\pi} \int_0^{2\pi}\! \mathrm{d}\chi_B \, 
	\int_0^\pi\! \mathrm{d} \theta_B \, \sin\theta_B \, 
	{\mathcal D}^{K}_{0 Q}(R_B)^\ast = \delta_{K 0} \delta_{Q 0} \, , 
\label{Miciso1}\\
	& \frac{1}{4\pi} \int_0^{2\pi}\! \mathrm{d}\chi_B \, 
	\int_0^\pi\! \mathrm{d} \theta_B \, \sin\theta_B \, 
	{\mathcal D}^{K'}_{Q Q'}(R_B) \, {\mathcal D}^{K''}_{Q Q''}(R_B)^\ast  
	= \frac{1}{2K' +1} \delta_{K' K''} \delta_{Q' Q''} \, , 
\label{Miciso2}
\end{align}
\end{subequations}
where for the second equation the Weyl's theorem has been used. 
Note that these same averages are performed on Eqs.~\eqref{ConvScheme1} 
or \eqref{ConvScheme2}, when calculating the line part of 
$\Delta {\mathcal E}^{0}_0$ with the Jacobi method.
It has to be observed that the only nonzero coefficient in the 
propagation matrix is now:
\begin{align}
	\mean{\eta_I(\nu,\vec{\Omega})} & = k_L \sum_{M_\ell M_u q} 
	\left(\begin{array}{c c c} 
		J_u & J_\ell & 1 \\
		-M_u & M_\ell & q
 	\end{array}\right)^2 \phi(\nu_{M_u M_\ell} - \nu) \; . 
\label{AbsIso}
\end{align}
Therefore, the four Stokes parameters will not be coupled in this case, and so 
the DELOPAR method shall not be required to solve the RT equation.  
From Eqs.~\eqref{EmisExpans}, \eqref{SourceLine}, and \eqref{SourLine}, and 
performing the field average described in this section, the following 
expression for the line emission coefficient, in the presence of an 
isotropic micro-structured magnetic field, is obtained:
\begin{align}
	\mean{\varepsilon_i(\nu,\vec{\Omega})^\ell} &  = k_L \sum_{K Q Q'} 
	\frac{1}{2 K + 1} {\mathcal T}^{K}_{Q}(i,\vec{\Omega})
	 \int\!\mathrm{d}\nu' \, (-1)^{Q} J^{K}_{-Q}(\nu') \, 
	R^{K' K'}_{Q'}(\nu',\nu,B) \notag \\
	& \; + k_L \frac{\epsilon'}{1 + \epsilon'} \Phi^{00}_0(J_\ell,J_u,\nu) 
	B_{\nu_0}(T) \, .
\label{EmisIso}
\end{align}

\noindent 
{\it b) Micro-structured field with fixed inclination and random azimuth}:\\
Fixing the inclination $\theta_B$ and averaging over the azimuth implies the 
evaluation of the following integrals:
\begin{subequations}
\label{Micran}
\begin{align} 
	\frac{1}{2\pi} \int_0^{2\pi} & \! \mathrm{d}\chi_B \, 
	{\mathcal D}^{K}_{0 Q}(R_B)^\ast = d^K_{00}(\theta_B) \delta_{Q 0} \, , 
\label{Micran1}\\
	\frac{1}{2\pi} \int_0^{2\pi} & \! \mathrm{d}\chi_B \, 
	{\mathcal D}^{K'}_{Q Q'}(R_B) \, {\mathcal D}^{K''}_{Q Q''}(R_B)^\ast  
	= \notag\\
	& (-1)^{Q-Q'} \sum_{\kappa} \, (2\kappa + 1) \left(
  	\begin{array}{c c c}
   		K' & K'' & \kappa \\
		Q & -Q & 0\\
	\end{array}\right) 
	\left(\begin{array}{c c c}
		K' & K'' & \kappa \\
		Q' & -Q'' & 0
	\end{array}\right) 
	d^\kappa_{0 0}(\theta_b) \; .
\label{Micran2}
\end{align}
\end{subequations}
The RT coefficients appearing in Eq.~\eqref{RTE}, 
as well as the corrections $\Delta {\mathcal E}^{0}_0$ that have to 
be calculated at each iteration, can be found following the same 
procedure as for the case a), 
but using the field averages shown in Eqs.~\eqref{Micran} 
instead.
\section{Illustrative results}
\label{sec_res}
In this section, we present a few illustrative applications of our RT code to 
resonance lines of diagnostic interest. 
Detailed investigations of specific spectral lines will be discussed in further 
publications.
All following calculations have been carried out in the semi-empirical solar 
atmospheric model C of \citet{Fontenla+93}. 
For the calculated linear polarization signal, we always take the reference 
direction for positive $Q$ perpendicular to the Z-axis of our reference system 
(the local vertical).
The RT problem is first solved using the RH code for the unpolarized case. 
The converged solution provided by this code is used as  initial guess for 
$J^0_0$ and $J^2_0$ for our RT code, which we have developed following the 
theoretical approach described in previous sections (hereafter; the 
Hanle-Zeeman code). 
The population of the lower level, as well as the continuum RT coefficients 
and scattering cross section have also been obtained from the RH code. 
In order to accurately calculate the populations of the lower level, in the
RH code we have considered a multi-level atomic model and we have considered
the impact of bound-free transition via photoionization and collisional 
ionization processes. 

For a given line, we estimate the formation height at a particular
frequency as the atmospheric height at which the corresponding optical depth
$\tau$ along the line-of-sight is unity.
At such height, we can estimate the fraction of coherent scattering processes 
(i.e., processes that are not perturbed by an elastic collision) through the 
so-called coherence fraction defined as: 
\begin{equation}
 \alpha = \frac{\Gamma_R + \Gamma_I}{ \Gamma_R + \Gamma_I + \Gamma_E}
\end{equation}

Our first application is for the Sr~{\sc i} line at $4607$~{\AA}.
This photospheric resonance line is produced by a transition with $J_u=1$ and 
$J_\ell = 0$, and it has a Hanle critical field $B_H = 23$~G.\footnote{The 
so-called Hanle critical field is the field strength characterizing the onset 
of the Hanle effect. It can be shown that 
\unexpanded{$B_{H} = 1.137\cdot10^{-7}\frac{\Gamma_R}{g_u}$.}} 
For the modeling of this line, we consider the depolarizing collisional 
rate $D^{(2)}$ given in \citet{Faurobert+95}.
As a second application, we consider the Sr~{\sc ii} line at $4078$~{\AA}, a 
resonance line forming in the low chromosphere, with $J_u = 3/2$ and 
$J_\ell = 1/2$.
The Hanle critical field of this line is 12~G, approximately. 
In the modeling of this line, we neglect the depolarizing effect of 
elastic collisions (i.e., we set the rate $D^{(2)}=0$).
For all the illustrative applications presented in this section, we consider 
the radiation emitted along a line-of-sight with $\mu = 0.1$, where $\mu$ is 
the cosine of the heliocentric angle.

\subsection{PRD calculation vs CRD limit}
\label{subsec_CRD}
\begin{figure*}[htp]
\centering
\includegraphics[width=\textwidth]{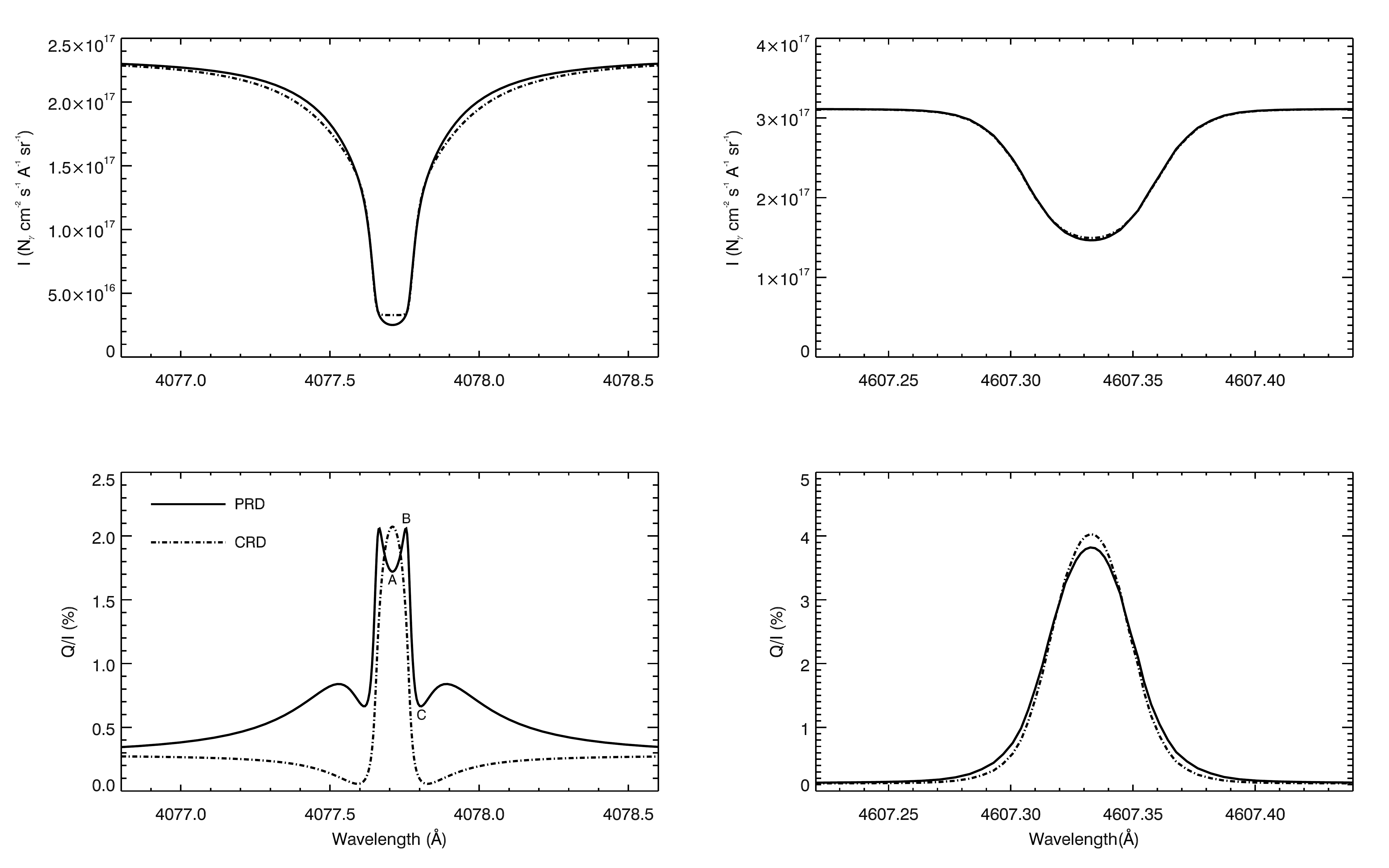}
\caption{Intensity (top row) and Stokes $Q/I$ (bottom row) profiles of 
the emergent radiation calculated for a line of sight with 
$\mu=0.1$ in the absence of a magnetic field, for the Sr~{\sc ii} line at 
$4078$~{\AA} (left column) and for the Sr~{\sc i} line at $4607$~{\AA} 
(right column), considering the FAL-C atmospheric model. 
In each figure, the results for the full PRD calculation (solid line) and the 
results obtained in the CRD limit (dash-dotted line) are plotted. The reference
direction of positive $Q/I$ is taken perpendicular to the vertical direction.}
\label{StokesFig1}
\end{figure*} 
The complex RT problem considered in this work becomes much simpler under 
the assumption of CRD, i.e., under the assumption that in a scattering process 
the frequencies of the incoming and outgoing photons are completely 
uncorrelated. 
Therefore, it is very important to clarify when this limit can be safely 
applied.
As mentioned before, the hypothesis of CRD is generally suitable for 
treating weak spectral lines, and it is a good approximation for modeling the 
core region of strong lines of the intensity spectrum. 
On the other hand, it turns out to be completely unsuitable for modeling the 
extended wings of strong resonance lines.
In light of this, we provide now a detailed comparison between the results 
obtained through full PRD calculation and in the limit of CRD.
In Fig~\ref{StokesFig1}, the emergent intensity and $Q/I$ profiles 
calculated in the absence of a magnetic field are shown, for the Sr~{\sc ii} 
$4078$~{\AA} line and for the Sr~{\sc i} $4607$~{\AA} line, comparing the full 
PRD calculation (solid line) to the calculation in the CRD limit (dash-dotted 
line).
This limit has been obtained by setting the elastic collision rate 
$\Gamma_E \rightarrow \infty$ while keeping all other parameters, including the 
$D^{(K)}$ depolarizing collision rates, as they were. 
In other words, we artificially modify the branching ratios so that all 
scattering processes occur through the $R_{\mbox{\sc \footnotesize III}}$ 
redistribution matrix. 
For both lines considered, the intensity is greater at line center when using 
the CRD approximation than when PRD is considered.
This can be qualitatively understood as follows. In the limit of CRD the 
scattered radiation at a given frequency depends on a weighted average of the 
incoming radiation over the whole line profile, and it thus takes into account 
that, for an absorption line, the intensity increases going from the core to 
the wings. 
On the contrary, in the limit of CS the scattered radiation is related only to 
the incoming radiation at that same frequency. PRD phenomena relax such 
coherency, relating the scattering radiation to the incoming one, but over an 
interval that is generally smaller than the one considered in the CRD case.
This same reasoning explains why in PRD the intensity is instead larger 
in the wings. The same behavior can be seen when comparing PRD and CRD 
calculations of the fractional linear polarization $Q/I$ profiles.

Fig.~\ref{StokesFig1} clearly shows that the difference between CRD and PRD 
calculations is, as expected, rather small in the Sr~{\sc i} 4607~{\AA} line. 
In particular, 
we expect that, once the 
Stokes profiles are smeared to properly account for the effects of the 
atmosphere's macroturbulent velocity and of the finite spectral resolution 
of a typical instrument, the differences between both calculations will be 
negligible. 
The results shown for this line are in agreement with those obtained
by \citet{Faurobert93}.
Although the Sr~{\sc i} line at 4607~{\AA} forms in the photosphere, 
where the density of neutral hydrogen (the main responsible for elastic 
collisions) is rather high, at the line center and for a line-of-sight with 
$\mu=0.1$, the coherence fraction $\alpha$ is 0.624, which means that 
$R_{\mbox{\sc \footnotesize II}}$ still represents the dominant contribution.
The good agreement between PRD and CRD calculations is due to the fact that 
this is medium/weak spectral line without extended wings outside the Doppler 
core. Indeed, it has to be recalled that 
when Doppler redistribution is taken into account, the emergent profiles 
produced by $R_{\mbox{\sc \footnotesize II}}$ and 
$R_{\mbox{\sc \footnotesize III}}$ in the line core are very similar 
\citep[see the discussion in][]{Thomas57}.
A detailed discussion of the behavior of the two redistribution functions 
depending on the optical thickness of a spectral line in the core and in the 
wings can be found in \citet{Faurobert87}.

The Sr~{\sc ii} $4078$~{\AA} line forms much higher in the atmosphere, where 
the density of perturbers is noticeably lower.
 Indeed, at the estimated formation heights for frequencies\footnote{See 
the lower left panel of Fig~\ref{StokesFig1}} A ($4077.7091$ \AA), 
B ($4077.7554$ \AA) and C ($4077.8064$ \AA) for a LOS with $\mu=0.1$, the 
coherence fractions are $0.998$, $0.996$ and $0.280$, respectively.
The contribution of $R_{\mbox{\sc \footnotesize II}}$ thus largely 
dominates in the core. In this region, however, the CRD limit still provides 
a rather good approximation since, as previously pointed out, 
$R_{\mbox{\sc \footnotesize II}}$ and $R_{\mbox{\sc \footnotesize III}}$ 
produce similar emergent profiles once Doppler redistribution is taken into 
account.
Unlike the Sr~{\sc i} $4607$~{\AA} line, 
this line presents very extended wings 
outside the Doppler core, where the optical thickness remains considerable.
In this region, the two redistribution functions have a different behavior, 
and they
give rise to emergent polarization profiles very different from one another.
It has to be noticed that the difference between the CRD and PRD $Q/I$ profiles 
remains significant in the near wings also at wavelengths where
the contribution of $R_{\mbox{\sc \footnotesize II}}$
is no longer the dominant one (see the coherence fraction at frequency point C).
The three peak structure found in Stokes $Q/I$, with a small dip in the central 
one, is characteristic of coherent scattering and cannot be found in the CRD 
limit. 
Notice also that for both lines considered in the figure, there is a non-zero 
$Q/I$ profile in the far wings which is produced by continuum scattering.
\subsection{Hanle-Zeeman calculation vs Hanle limit}
\label{subsec_HZ1}
 When the magnetic field is not too strong (e.g., outside sunspots), the 
splitting of the magnetic sublevels due to the Zeeman effect is 
generally much smaller than the Doppler width of the line. 
Under this circumstance, it is customary to neglect the Zeeman splitting 
of the energy levels in the emission and absorption profiles, as this leads 
to a significant simplification of the problem. 
Under this assumption, which is generally referred to as the weak-field 
approximation, the Zeeman effect is negected by definition.
\begin{figure*}[htp]
\centering
\includegraphics[width=\textwidth]{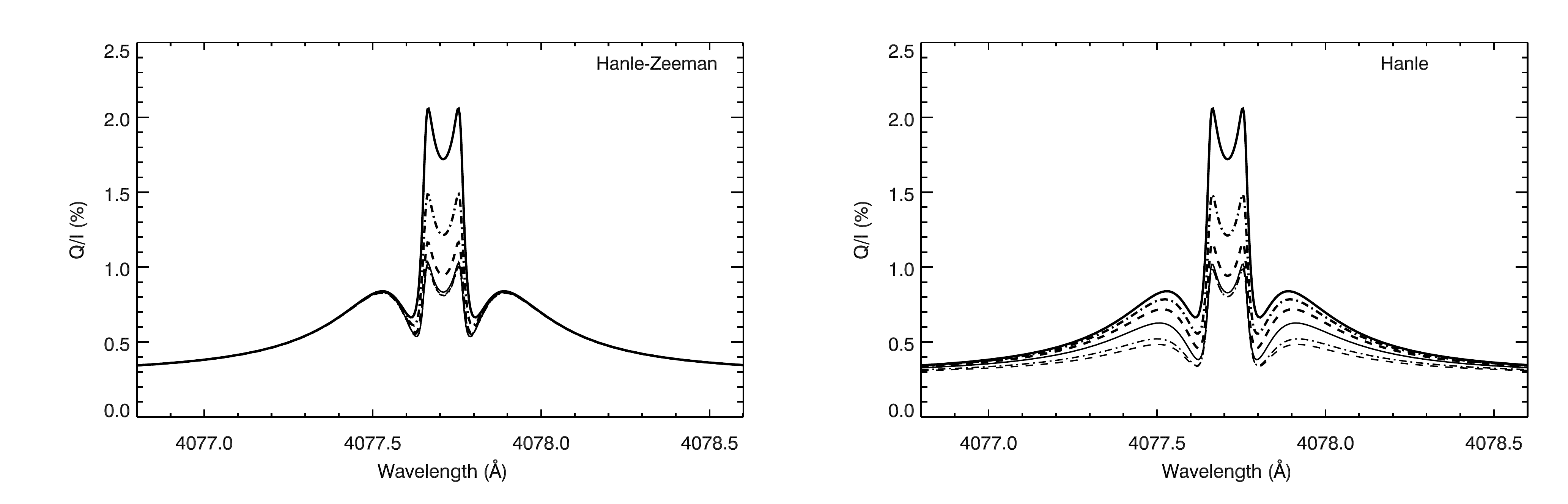}
\caption{Stokes Q/I for the emergent radiation at line of sight $\mu=0.1$ for the Sr {\sc ii} line at $4078$ \AA\ in the presence of a deterministic magnetic field with $\theta_B = \pi/2$ and $\chi_B = \pi/2$ for various field strengths. Left panel: Full calculation. Right panel: Calculation in the weak field approximation (Hanle limit). For both cases the curves correspond to the following field strengths: 0 G (thick solid line), 5 G (thick dash-dotted line), 10 G (thick dashed line), 20 G (thin solid line), 50 G (thin dash-dotted line), 100 G (thin dashed line)}
 \label{StokesFig2}
\end{figure*} 
 In this section, we analyze 
the impact of using this approximation on the calculated 
scattering polarization profiles, when PRD phenomena are taken into 
account.
In Fig.~\ref{StokesFig2} a comparison of the full calculation and the 
calculation using the weak field approximation for $Q/I$ is shown for a 
horizontal and transverse magnetic field, for the Sr~{\sc ii} line at 
$4078$~\AA . Given that the magnetic field is transverse, 
there is no Hanle rotation, and for this reason $U/I$ is
not shown in this case. For the same reason, $\rho_V$ is zero in this case
(see the following section). 
At line center, the polarization in both cases is identical up to 50 G, and 
there is a very small discrepancy at 100 G due to a small contribution from 
the Zeeman effect.
 In the wings, however, a clear magnetic depolarization can be 
observed when the weak field approximation is considered.
This is an artificial effect caused by the fact that we are neglecting the 
Zeeman splitting in the emission profile, and so the Hanle depolarization 
factor no longer cancels the magnetic field dependence in the wings (see 
Sect.~10.4 of LL04 for a more detailed discussion). 
It is important to note that, though the weak field approximation is reliable 
in the line core (as long as the conditions for its validity are met), when 
there are extended polarization signals in the wing - as often occurs when 
considering the effects of PRD on modeling strong resonant lines - one must 
be cautious in considering the effects of the magnetic field with this 
approximation. 
The results for stronger fields are not shown because when Zeeman splitting 
is neglected there is no change in the polarization profiles once we have 
reached Hanle saturation. 
\subsection{The impact of magneto-optical effects}
\label{subsec_HF0}
We consider now the emergent $Q/I$ and $U/I$ profiles of the Sr~{\sc ii} 
line at 4078~{\AA}, calculated for a LOS with $\mu=0.1$, in the presence of a 
magnetic field with $\theta_B = \pi/2$ and $\chi_B = 0$ (i.e., a magnetic 
field almost longitudinal for the considered LOS).
\begin{figure*}[htp]
\centering
\includegraphics[width=\textwidth]{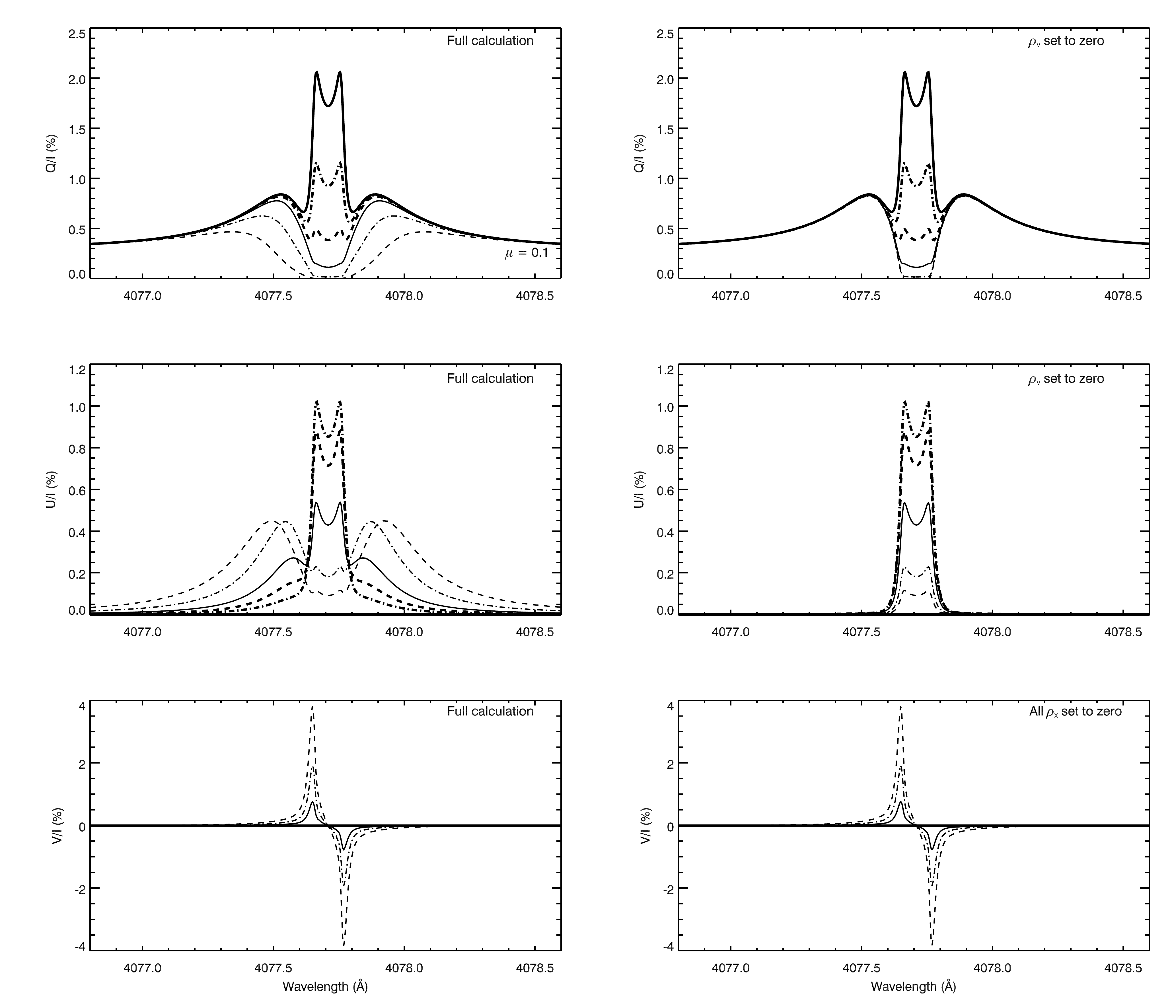}
\caption{ Fractional polarization $Q/I$, $U/I$ and $V/I$ profiles of the 
radiation emergent at $\mu = 0.1$ for the Sr~{\sc ii} line at $4078$~{\AA} 
in the presence of a deterministic magnetic field with $\theta_B = \pi/2$ 
and $\chi_B = 0$, for various field strengths. 
In the left panels the full calculations for Stokes $Q/I$, $U/I$ and $V/I$ are presented,
while in the right panels calculations are shown in which the magneto-optical 
term $\rho_V$ of the propagation matrix has been neglected for $Q/I$ and $U/I$,
and all $\rho_X$ terms have been neglected for $V/I$. 
The curves represent the same field strengths as in Fig.~\ref{StokesFig2}. However, to facilitate the visibility of the curves, for $V/I$
the 5 G and 10 G cases are not shown.}
\label{StokesFig3}
\end{figure*} 
 As can be observed in the left panels of Fig.~\ref{StokesFig3}, 
weak magnetic fields, even considerably below the Hanle critical field, 
produce a clear depolarization of the wings of the calculated $Q/I$ profiles,
and give rise to positive lobes in the wings of the $U/I$ profiles.
These effects become larger and extend further out into the line wings 
as the field strength grows.
At first glance, this magnetic sensitivity of the line wings may appear 
surprising, as we know that the Hanle effect vanishes in the wings, and the 
contribution of the Zeeman effect to $\varepsilon_Q$ and $\varepsilon_U$ 
is expected to be negligible for these field strengths.
Indeed, this magnetic sensitivity has nothing to do with such   
effects, but it is a magneto-optical effect, due to the term $\rho_V$.  
As it is clear from Eq.~\eqref{RTE}, this term couples $Q$ and $U$, and produces
a rotation of the plane of linear polarization as the radiation propagates 
through the atmospheric material (Faraday rotation).
 We recall that in the absence of lower level polarization (as in our 
case), $\rho_V$ is zero unless a magnetic field with a longitudinal component 
is present. The coefficient $\rho_V$ is proportional to the (antisymmetric) 
Faraday-Voigt profile: it is thus zero at the line center, but has very 
extended wings, where it can be even larger than the absorption coefficient
$\eta_I$. 
This explains why the magnetic sensitivity shown by our calculations is 
so clear in the line wings, while it disappears in the core of the line.
Our physical interpretation of this effect is clearly supported
by the results obtained by neglecting the magneto-optical 
effects caused by the $\rho_V$ coefficient appearing in the propagation matrix of Eq. (1).
As shown in the right panels of Fig.~\ref{StokesFig3}, when $\rho_V$ is set 
to zero the line core signal shows exactly the same behavior as when this 
term is taken into account, while in the wings there is no depolarization at 
all in $Q/I$, and no signal appears in $U/I$.
It should be observed that in our calculations we have considered an 
almost longitudinal field, constant throughout the whole atmosphere, a 
scenario that is particularly suitable for the illustration of this physical 
effect. It is also worth mentioning that in the Hanle-Zeeman calculation
a Stokes $V/I$ profile is produced, and it can be seen that it is not affected
by the $\rho_V$ (as can be easily deduced from Eq.~\ref{RTE}). Moreover, for this 
geometry and field strength, $\rho_Q$ and $\rho_U$ are too weak to produce any
appreciable change in the emerging circular polarization (see Fig.~\ref{StokesFig3}).

The magnetic
sensitivity of the linear polarization in the wings of the lines, will
be discussed in more detail in future publications focused on 
 several chromospheric spectral lines.
\section{Conclusions}
In order to correctly model the scattering polarization signals 
of strong resonance lines in an optically thick plasma, in the presence of 
magnetic fields of arbitrary intensity and orientation, it is necessary 
to solve a complex non-LTE radiative transfer problem,
taking into account the joint action of the Hanle and Zeeman effects, 
as well as the impact of PRD phenomena.
In this work, we have considered
the theoretical approach of \cite{Bommier97a,Bommier97b}, which is 
capable of accounting for all these physical ingredients, 
and we have developed and applied a series of 
numerical methods required for the efficient and accurate solution of the equations involved.

The resulting radiative transfer code provides a new tool for solar and stellar spectropolarimetry. 
It considers a two-level atomic model with an unpolarized and infinitely 
sharp lower level, which is suitable for investigating the magnetic sensitivity of 
several resonance lines of diagnostic interest such as Sr~{\sc ii} $4078$~{\AA}, Sr~{\sc i} $4607$~{\AA}, or Ca~{\sc i} 
$4227$~{\AA}.

The above-mentioned theoretical approach is based on the 
redistribution matrix formalism. 
The total redistribution matrix is given by a linear 
combination of two terms: one describing coherent scattering processes 
(${\mathcal R}_{\mbox{\sc \footnotesize II}}$) and 
another describing scattering processes in the limit of complete 
frequency redistribution (${\mathcal R}_{\mbox{\sc \footnotesize III}}$). 
We have started from the expressions provided in \cite{Bommier97b}, 
valid in the atomic frame, taking the quantization axis directed along the 
magnetic field. We have shown how to rotate them in a reference system with the quantization 
axis directed along an arbitrary direction, and how to transform them
from the atomic rest frame into the frame of the observer. 
The expressions corresponding to the case in which a magnetic field 
that changes its direction over scales smaller than the line 
photon's mean free path have also been studied.

We have presented illustrative results for the Sr~{\sc i} photospheric line at 4607~{\AA} and 
for the Sr~{\sc ii} chromospheric line at 4078~{\AA}, and in forthcoming 
publications we will describe in detail other interesting applications to the $k$ line of Mg~{\sc ii} 
at 2795 \AA\ and to the Ca~{\sc i} line at 4227 \AA.  The main results are the following:
\begin{itemize}
\item{{\it The impact of PRD phenomena}. Calculations accounting for the effects of PRD have been compared to those in the 
CRD limit, in order to quantitatively evaluate the suitability of this approximation. In photospheric lines without significant wings
such as Sr~{\sc i} 4607 \AA , we can confirm that the CRD limit 
is a very good approximation for modeling the intensity and scattering polarization. Nevertheless for strong
chromospheric lines, with extended wings outside the Doppler core, 
such as Sr~{\sc ii} 4078 \AA\, the impact of PRD phenomena is very significant, 
especially in the near wings.
The resulting scattering polarization profiles show 
extended wings and complex multi-peak structures. Such profiles cannot be found in the limit of CRD, which however keeps
 representing a quite good approximation for modeling the line-center amplitude of both intensity and scattering
 polarization signals.
While in the atomic reference frame coherent scattering effects play an important role also in the
line center, in the observer's frame the effects of Doppler redistribution cause the CRD
approximation to be suitable to estimate the polarization at the line center. But in the near wings the effects of PRD 
need to be taken into account, for both the intensity and the emergent scattering polarization.}

\item{{\it The weak-field approximation in the general PRD case}. 
In another application to the Sr {\sc ii} 4078 \AA\ line we compared the results obtained when applying the weak field 
approximation with the results of our Hanle-Zeeman calculation. While in the line core the resulting scattering polarization 
signals agree, in the wings we find that the results for the weak-field approximation become inaccurate, since artificial signals
 are found when neglecting the Zeeman splitting in the absorption and emission profiles.}

\item{{\it Magneto-optical effects in the general PRD case}. Furthermore, we have found that in strong resonance lines for 
which PRD effects produce sizable $Q/I$ wing signals, such as that of Sr {\sc ii} at 4078 \AA, 
a novel physical mechanism operates that creates $U/I$ wing signals and introduces a very interesting 
magnetic sensitivity in the wings of the $Q/I$ and $U/I$ profiles. This magnetic sensitivity has nothing to do with the Hanle effect, nor with the Zeeman effect in emission. Instead, we conclude that it is caused by magneto-optical effects; in particular, by the coupling between Stokes $Q$ and $U$ due to the $\rho_V$ term of the propagation matrix as the radiation propagates through the 
magnetized solar atmosphere.}
\end{itemize}

\acknowledgements

We would like to thank the anonymous referee for insightful and 
helpful comments.
We are also grateful to Tanaus\'u del Pino Alem\'an (HAO) for several 
scientific discussions during the development of this work that were very 
helpful to refine the numerical method presented here for solving the PRD 
radiative transfer problem for arbitrary magnetic fields. 
Likewise, we are grateful to Egidio Landi Degl'Innocenti (University of 
Firenze) for illuminating discussions on the applied theory of spectral line 
polarization. 
Financial support by the Spanish Ministry of Economy and Competitiveness 
through projects \mbox{AYA2014-55078-P} and \mbox{AYA2014-60476-P} is 
gratefully acknowledged. E. Alsina Ballester also wishes to acknowledge the 
Fundaci\'on La Caixa for financing his Ph.D. grant.


\end{document}